 \documentstyle[11pt,aaspp4]{article}
\def\nhi{N(\textsc{H~i})}
\def\nciv{N(\textsc{C~iv})}
\def\cm2{cm$^{-2}$}
\def\kms{km s$^{-1}$}
\def\lya{Ly$\alpha$\ }
\def\civ{\textsc{C~iv}\ }
\def\nrange{$10^{13.5}<{\rm N(\textsc{H~i})}<10^{14}$ cm$^{-2}$}
\begin{document}

\title{The Metal Contents of Very Low Column Density Lyman-alpha Clouds:
Implications for the Origin of Heavy Elements in the Intergalactic Medium$^1$}
\author{Limin Lu$^{2}$, Wallace L. W. Sargent$^2$, Thomas A. Barlow$^3$,
and Michael Rauch$^{2,4}$}
\altaffiltext{1}{Based on observations obtained at the W. M. Keck
observatory, which is jointly operated by the California Institute of
Technology and the University of California.}
\altaffiltext{2}{California Institute of Technology, 105-24, Pasadena, 
CA 91125}
\altaffiltext{3}{California Institute of Technology, 100-22, Pasadena,
 CA 91125}
\altaffiltext{4}{Hubble Fellow}

\begin{abstract}
   We investigate the metal contents of \lya clouds at $2.2<z<3.6$ with 
\nrange\ using high resolution, high S/N spectra of 9 quasars obtained
 with the
10m Keck I telescope. Previous investigations of similar nature were
limited to \lya clouds with \nhi$>10^{14.5}$ \cm2 for which a mean
metallicity of [C/H]$\simeq -2.5$ was deduced. It has been suggested
that the metals seen in these clouds may have been produced by a generation 
of Population III stars occurred at a much earlier epoch before the 
formation of quasars and normal galaxies. 
We shift the quasar spectra into the rest frame
of each \lya cloud  with \nrange and then average
the rest-frame spectra in the spectral region encompassing the \civ
$\lambda\lambda$1548.20, 1550.77 absorption lines in order to vastly
improve the S/N of the final composite spectrum. 

After eliminating a small number of \lya lines that are estimated to 
have \nrange which individually show detectable \civ absorption,
the remaining $\sim 300$ lines whose corresponding
\civ $\lambda$1548.20 regions are clean are divided into various
samples to form composite spectra. No significant \civ absorption is
detected in any of the composite spectra we investigated. 
We derive an upper limit of $<0.088$
m\AA\ (95.5\% confidence limit) for the equivalent width of the
\civ $\lambda$1548.20 line, corresponding to [C/H]$<-3.5$
using the cosmological simulations of Rauch, Haehnelt,
\& Steinmetz (1997) to 
infer the ionization correction; the same simulation results have
been used to deduce a mean metallicity of [C/H]$\simeq -2.5$ for \lya clouds
with \nhi$>10^{14.5}$ \cm2. 

The mean metallicities of \lya clouds with \nrange, [C/H]$<-3.5$, 
is a factor of 10 lower than that inferred for \lya clouds with
\nhi$>10^{14.5}$ \cm2. This suggests that a sharp drop in the 
metallicity level
of the intergalactic gas sets in at $10^{14}<$\nhi$<10^{14.5}$ \cm2.
The result rules out the suggestion that a generation of Pop III stars
could have polluted the entire universe to a (nearly) uniform metallicity
level of [C/H]$\simeq -2.5$.
Cosmological simulations involving  gas hydrodynamics indicate that
\lya clouds with \nhi$>10^{14.5}$ \cm2 mostly occur in the filamentary  
gaseous regions surrounding and connecting collapsed objects, while those
with \nhi$<10^{14}$ \cm2 are preferentially found in void regions further 
away from collapsed structures. These results, coupled with the simulation
results of Ostriker \& Gnedin (1996) and Gnedin \& Ostriker (1997)
for Pop III star formation and enrichment,
strongly suggest that most of the heavy elements in \lya clouds 
with \nhi$>10^{14.5}$ \cm2 (i.e., gas in the filaments) were probably
produced {\it in situ} by Pop II stars, in the sense that 
they were either made by stars within
the clouds or were ejected from nearby star-forming galaxies.
Within this context, clouds with \nhi$<10^{14}$ \cm2 (i.e., gas in the void
regions) may only have experienced pollution from Pop III stars.

We derive constraints on the metallicities of the small number of \lya
clouds with \nrange that individually show detectable \civ absorption
and discuss their possible nature.

\end{abstract}


\section{INTRODUCTION}

   The \lya forest clouds seen in spectra of quasars, which are 
crudely defined to have \nhi$<10^{16}-10^{17}$ \cm2, were
thought initially to contain primordial gas since no metal absorption lines
were detected at their redshifts (Sargent et al 1980). 
However, moderate-resolution spectra taken with much higher S/N ratios
(Meyer \& York 1987) provided the first evidence that some of
these clouds, especially those with relative large equivalent width,
 do show weak \civ absorption. By shifting to the rest frame
and averaging  the \civ regions of hundreds of \lya lines which did not
show \civ absorption individually, Lu (1991) provided a probably detection
of \civ absorption associated with these clouds at 
the $w_r(1548)\sim 7$ m\AA\ level. Subsequent
high resolution, high S/N spectra taken with  the 10m Keck telescopes (Cowie
et al 1995; Tytler et al 1995; Songaila \& Cowie 1996) have clearly
revealed (mostly) weak \civ absorption associated with individual \lya clouds
at $z\sim 3$ with \nhi\ as low as $10^{14.5}$ \cm2. Specifically, 
it was found that essentially all clouds 
with \nhi$\geq 10^{15}$ \cm2 show \civ absorption,
and that 50-75\% of the \lya clouds with \nhi$\geq 10^{14.5}$ \cm2
show \civ absorption. 

   The presence of \civ absorption in \lya clouds with \nhi$> 10^{14.5}$ \cm2
indicates that these clouds have been enriched in heavy elements
at redshifts as high as $z>3.6$.  Recent cosmological 
simulations of structure formation involving gas processes (hydrodynamics) 
suggest that the \lya lines  at $z>2$ represent low density gas
in-between collapsed objects and are closely related to the formation of
galaxies, clusters, voids, and other structures in the universe (Cen et al
1994; Petitjean, Mucket, \& Kates 1995; Zhang, Anninos, \& Norman 1995; 
Hernquist et al 1996; Miralda-Escude et al 1996). 
An important issue is whether the metals seen in these intergalactic clouds
were produced locally (i.e., produced
by stars formed within the clouds, or 
ejected from nearby star-forming galaxies)
or were produced at a much earlier epoch in a wide-spread fashion 
(i.e., by Population III stars; Ostriker \& Gendin 1996).

If the metals in \lya clouds were produced {\it in situ}, one may expect 
that clouds in regions of space 
sufficiently far away from star-forming objects should be
metal-free. Whereas in the case of Pop III star enrichment, most of the 
universe may have been polluted to some 
degree and few clouds may be entirely metal-free. 
While it is not yet possible to directly see the spatial distributions of
star-forming objects and \lya clouds at high redshift,
some insight is provided by cosmological
simulations. These simulations indicate that clouds with
$10^{14.5}<$\nhi$<10^{17}$ \cm2 account for 40-60 \% of the total 
baryons in the universe at $z=2-4$ (Miralda-Escude et al 1996),
and that such clouds preferentially
occupy the filamentary gaseous regions surrounding and connecting collapsed 
structures (i.e., galaxies). The simulations also indicate that clouds
with \nhi$<10^{14}$ \cm2 predominantly occupy the void regions. These
results lead to the idea that one can perhaps obtain clues about the origin
of the metals by investigating the metal contents of the very low column 
density (i.e., \nhi$<10^{14}$ \cm2) clouds: 
the presence of any significant amount of metals in these clouds
would favor the Pop III idea for the reasons discussed above.

In this paper we present the results of a search for \civ absorption 
associated with  \lya clouds with $10^{13.5}<$\nhi$<10^{14}$ \cm2, using
high resolution, high S/N quasar spectra obtained with the 10m Keck I 
telescope and the High Resolution Spectrometer (HIRES; Vogt 1992).
Given the low \nhi\ of the clouds,
it would be practically impossible to obtain spectra of quasars
with sufficient S/N to detect the \civ absorption in
individual clouds even if the N(\civ)/\nhi\ ratio remains
the same as that in clouds with higher \nhi\
(S/N at least a factor of 5 better than the current
best spectrum would be required). The problem is compounded 
by the fact that the N(\civ)/\nhi\ ratio is predicted to drop with
decreasing \nhi\ at \nhi$<10^{14}$ \cm2 as the clouds become more ionized
(e.g. Rauch, Haehnelt, \& Steinmetz 1997). 
Accordingly, the technique we adopt is
the composite spectrum method pioneered by Norris, Hartwick, \& 
Peterson (1983).
Essentially, one shifts a quasar spectrum into the
rest frame of each \lya absorption line and then averages
the rest-frame spectra over
the spectral region encompassing the \civ $\lambda\lambda$1548.20, 1550.77
lines to form a composite spectrum.
The S/N in the composite spectrum increases roughly as the
square root of the number of lines being averaged, making this a very
effective way to achieve ultra-high S/N without using unrealistically
large amounts of telescope time. 
The end result is then an average \civ absorption 
associated with an ensemble of \lya clouds. This technique has
been applied previously to \lya clouds to search for \civ absorption
by Williger et al (1989), Lu (1991),
Tytler \& Fan (1994), and Barlow \& Tytler (1998), 
and to metal systems to search for O VI absorption
by Lu \& Savage (1993). Incidentally, this averaging process
automatically suppresses any residual noise features in the spectra caused by
imperfect flatfielding or other artifacts and 
is ideal for achieving ultra-high S/N in the averaged spectra.

    In section 2 we describe the spectra used in the analysis and the
procedure for selecting \lya lines with the desired column densities
(i.e., \nrange). The composite spectra for various samples of \lya clouds
are discussed in section 3. No \civ absorption is detected in any of the
composite spectra, thus yielding only upper limits to the metallicity of
the clouds. In section 4 we discuss the implications of our results for
the nature of the metal enrichment in the \lya clouds. We briefly summarize
our main conclusions in section 5.

\section{THE Ly$\alpha$ LINE SAMPLE}

\subsection{Selection of Ly$\alpha$ Lines}

   The list of 9 quasars whose spectra are used in the analyses
is given in Table 1. These objects are selected because their spectra
have the highest S/N among all $z_{em}>2.5$ quasars observed by the
current authors with the Keck HIRES. The objects have emission redshifts
between 2.58 to 3.63.  The integration times range from 4 to 11 hours. 
The data were reduced using the EE package developed by T. Barlow.
The typical S/N per resolution element (6.6 \kms) in the \civ region
for each spectrum is given in the 4th column of Table 1.

   Our goal is to determine the metallicity of \lya clouds with \nhi\
in the range of $10^{13.5}$ to $10^{14}$ \cm2 via the \civ $\lambda\lambda$
1548.20, 1550.77 absorption. Such clouds could be
selected by fitting Voigt profiles to all the \lya lines in the spectra
as is commonly done.
Since such profile fitting has not been done to all of the 
spectra listed in Table 1
and indeed would be quite time consuming to carry out, we instead
selected \lya
lines in the desired column density range based on the observed depth
of the lines at the line center. 
Previous Voigt profile fitting to \lya forest
absorption lines obtained with the Keck HIRES indicate  that the Doppler 
width distribution of the \lya lines at $2<z<4$ is sharply peaked
at $b\simeq 25$ km/s with the vast majority of lines having $b$ between
20 and 50 \kms (Hu et al 1995; Lu et al 1996a; Kirkman \& Tytler 1997).
For a mean $b=25$ \kms, the central optical depths
corresponding to the limiting \nhi\ of $10^{13.5}$ and $10^{14}$ \cm2
are 0.96 and 3.03. The corresponding absorption lines should have
a residual flux between 0.384 and 0.048 at the line center for unity 
continuum
(note that the lines are completely resolved at the HIRES resolution
of FWHM=6.6 \kms for the assumed $b$ value).
We therefore selected \lya clouds with \nrange by
visually selecting \lya lines with residual flux between
0.384 and 0.048 after normalizing the spectrum to unity 
continuum.\footnote{Some of the spectra contain damped \lya absorption
lines (cf. Wolfe 1988). In such cases, the adopted continua take into 
account the wings of the damped \lya absorption.}
In order to avoid biasing the line sample, only spectral regions which
have sufficiently high S/N to securely identify absorption lines with
residual flux of 0.048 were used. The spectral regions actually
used to select \lya lines are given in column 5 of Table 1.
The lower wavelength limit is determined either by the above S/N
requirement or by the onset of Ly$\beta$ emission of the quasar (we do
not use the spectral region shortward of Ly$\beta$ emission in order
to avoid mis-identifying Ly$\beta$ and higher order Lyman
lines with \lya). The upper 
wavelength limit is determined by the \lya emission.

   In many cases \lya lines with the desired optical depth (i.e., 
residual flux between 0.384 and 0.048) are blended with other lines.
If the blending is severe enough, two problems arise: 
(1) the blending makes
it difficult to determine an accurate redshift for the absorption line;
(2) the blending could lead to a significant overestimate of the \textsc{H~i}
column density of the absorption lines.
To avoid these problems, we excluded from the sample
\lya lines within 50 \kms of other ``strong'' lines  (\lya or 
something else) that would have otherwise been included in the sample.
A ``strong'' line is defined here to have residual flux $<0.384$
(i.e., \nhi$>10^{13.5}$ \cm2 if it is a \lya). 
The choice of 50 \kms is somewhat 
arbitrary and is based on the observation that \lya lines satisfying
the above criterion do appear to have well-defined redshifts
and largely fall into the desired column density range. It is important to
emphasize that, while  it is critical for the line sample to be unbiased with
respect to the column density distribution so that the average
\nhi\ of the lines comes out to be what is expected (see below), 
it is not necessary for the line sample to be complete.

   Most of the metal lines occurring in the \lya forest are easy to 
recognize owing to their narrow line width (see, for example,
Kirkman \& Tytler 1997). Relatively
broad metal lines that are strong enough to be mistaken as \lya lines
in the desired column density range are rare
and are also easy to identify 
because they are usually associated with prominent metal systems.
  Spectral regions
that are severely affected by metal absorption lines were not used
to search for \lya lines. 
There are a few cases where the presumed \lya lines have estimated
width of $15<b<18$ \kms, which could be \lya lines but could also
be unidentified metal lines; these are assumed
to be \lya lines. Since the number of such lines is very small
($<5$\%), their effect on the final composite spectrum and on the
interpretation should be negligible.

    The selection criteria we employed will occasionally
include some \lya lines with \nhi\ outside the range of 
$10^{13.5}-10^{14}$ \cm2 and exclude some other \lya lines within
the desired column density range if the actual Doppler widths of
the lines are significantly different from the assumed mean
of 25 \kms. For example, a line with $b=15$ \kms and residual 
flux 0.384 (i.e., barely strong enough
to be included in the sample if it had $b=25$ \kms) would have
\nhi$=10^{13.28}$ \cm2 but would still be included in the sample.
On the other hand, a line with $b=50$ \kms and residual flux
0.3 would have a \nhi=$10^{13.80}$ \cm2 and would be
excluded from the sample. However, because of the sharpness
of the Doppler distribution of the \lya clouds, and because of
the range of column densities considered, 
the vast majority of lines in the sample
should have the desired \nhi\ values. The only major effect of
the above ``shuffling'' is
to broaden the column density distribution
of the selected lines. The mean \nhi\ of the sample should be
largely unaffected.

   Kirkman \& Tytler (1997) have fitted Voigt profiles to all 
\lya lines in the HIRES spectrum of Q1946+7658 which they took
independently. It is therefore possible to examine the actual
\nhi\ distribution of the \lya lines in the Q1946+7658 spectrum
that are included in our sample based on the above selection criteria
using the results of Kirkman \& Tytler.
Figure 1 shows this \nhi\ distribution for the 47 systems which
we have selected based on their optical depth. 
As expected, some of the
systems (9/47 or 19\%) have \nhi\ outside the desired range, and the
extremes are \nhi$=10^{13.26}$ and $10^{14.10}$ \cm2. The average
\nhi\ of the 47  \lya lines included in the sample for Q1946+7658
is $10^{13.783}$ \cm2  based on the profile fitting results. The expected
mean column density from the column density distribution,
$f(N)\propto N^{-1.5}$ (Hu et al 1995; Lu et al 1996a; Kirkman \& 
Tytler 1997), is 
$10^{13.750}$ \cm2. Note that the mean \nhi\ is extremely robust.
For example, if the power law index is 1.4 rather than
1.5 as assumed, the expected mean \nhi\ becomes $10^{13.755}$. The 
robustness of the mean \nhi\ has to do with the fact that the
total column density range considered here spans only a factor of 3.
The true mean \nhi\ of any given sample will fluctuate according
to the sample size. Most of the samples we consider
here (Table 4) contain between 126 and 282 lines.
Monte Carlo simulations indicate that the 1$\sigma$ dispersion in
the mean \nhi\ for samples of such sizes ranges between 0.006 and 0.013
dex for the assumed \nhi\ distribution. The 1$\sigma$ dispersion in the mean
\nhi\ for the sample with 27 lines is 0.028 dex, and that for the sample
containing 12 lines is 0.04 dex.  These uncertainties are very small. 
The uncertainties associated with the mean \nhi\ of the actual
samples will be somewhat larger than what the simulations indicate
because of the ``shuffling'' effect discussed above. Even so, the
uncertainties associated with the mean \nhi\ should be small
compared to other uncertainties 
involved in the analysis (e.g., the ionization correction).
We therefore believe that the simple 
selection procedure employed above does form a sample of \lya
lines with the vast majority of them having the desired range in \nhi, 
and that the average \nhi\ of the sample is unbiased and  can be 
estimated reliably from the known column density distribution.
Finally, we note that while line blending and blanketing effects can bias
the observed \nhi\ distribution of \lya clouds, such biases are apprently
negligible at the column densities we consider here (see Kirkman \&
Tytler 1997, for example).

\subsection{Individual Ly$\alpha$ Systems with C IV Absorption}

   Before we formed the composite spectrum, we visually 
examined the \civ region of individual \lya systems in the sample
to see if any of them have  detectable \civ absorption in
the observed spectra. Sixteen such systems were found, 6 of which
form 3 close pairs ($\Delta v<50$ \kms) and may be physically
related. Three of the 16 systems are marginal in the sense that
only weak ($4-6\sigma$) \civ $\lambda$1548 absorption is detected
and the corresponding $\lambda$1550 absorption is not apparent.
We show in Figure 2 these \lya lines and the
corresponding \civ absorption. The properties of these systems
are summarized in Table 3 and are discussed in more detail in section 4.2. 
These systems will be excluded from the composite spectrum analyses.

Nine of the 16 systems with detected \civ\ occur within 3,000 \kms
of the \lya emission redshift of the quasars. It has been known
for some time
that there is in general an excess of metal absorption systems near
$z_{em}$ of quasars, which are likely associated with material ejected
from the quasar nuclei or in the host galaxies and their environment
(cf. Foltz et al 1988). We therefore classify these 9 systems
as ``associated'' systems in Table 3.
The remaining 7 systems are more than 10,000 \kms away from the
\lya emission redshift of the background quasars; these are
classified as ``intervening'' in Table 3.

Most of the systems listed in Table 3
 show no detectable Si \textsc{iv} absorption
except for the two pairs of associated systems at $z=3.62$ toward
Q 1422+2309 and at $z=2.56$ toward Q 2343+1232, which happen to
have (perhaps not by accident) the strongest \civ\ absorption
among all  systems given in Table 3. The $z=3.45129$
system toward Q 1422+2309 has a marginally significant 
Si \textsc{iv} $\lambda$1393 absorption without obvious
$\lambda$1402 absorption.

Of the 7 intervening systems, 4 are within 500 \kms of other
\civ absorption systems, while another one is within 800 \kms
of other \civ systems (see figure 2). 
Interestingly, the 2 systems that do not
have other \civ absorption systems within 1000 \kms, the
$z=3.31822$ system toward Q1422+2309 and the $z=2.50221$ system
toward Q2343+1232, are both
marginal in their \civ detection. These results suggest a possible
trend that low \nhi\ \lya clouds are more likely to show metal
absorption when they are close to other metal systems. This possibility
is explored later in section 3.
The 2 marginal systems that do not have other metal systems within
1000 \kms could be false identifications.

\section{COMPOSITE SPECTRA AND MEASUREMENTS}

\subsection{Composite Spectrum}

   Before we formed the composite spectrum, we visually inspected the
spectral region of each quasar between the \lya and \civ emission
lines and masked out any features (absorption lines, apparent
emission features due to cosmic rays, features due to poor subtraction
of night sky lines or telluric absorption lines, etc) that deviate
significantly from the continuum. This essentially removed all features
with significance above $\sim 2\sigma$ level (as judged by eye). 
We will refer to these masked-out 
regions as ``bad'' regions or regions containing
``bad pixels'', and refer to the resulting spectra as ``cleaned spectra''.
The bad regions are excluded from entering into the composite spectra.

   We then examined the \civ region corresponding to each
\lya line in the sample in the cleaned spectra, and assign a flag to
each system according to the ``cleanness'' of its \civ region: ``0''
for systems whose \civ $\lambda\lambda$1548.20, 1550.77 regions are both
clean, ``1'' for systems whose $\lambda$1550.77 region is clean but
whose $\lambda$1548.20 region contains bad pixels, ``2'' for systems 
whose $\lambda$1548.20 region is clean but whose $\lambda$1550.77 region
contains bad pixels, and ``3'' for systems that contain bad pixels in 
both the $\lambda\lambda$1548.20, 1550.77 regions. In the above,
``$\lambda$1548.20 region'' refers to the spectral region within
$\pm 30$ \kms of the expected \civ $\lambda$1548.20 line center; 
likewise for $\lambda$1550.77. 

  We then formed composite spectra by shifting the spectrum of
each system with certain chosen flag(s) (see section 3.3) to 
the rest frame of the \lya 
line, rebinning the rest-frame spectrum onto a common velocity scale of bin
size of 3.3 \kms (half the nominal resolution of the
HIRES spectra), and averaging all the rest-frame spectra in
the \civ region weighted by the
square of S/N at each bin. We also tried weighting each spectrum by the
square of the mean S/N within $\pm 30$ \kms of the \civ $\lambda$1548
line, with negligible change in the results.

  To estimate the equivalent width of the \civ absorption 
in a composite spectrum, a continuum was first fitted to the 
composite spectrum using cubic splines after excluding the
$\pm30$ \kms region centered on the $\lambda$1548.20
position. We then measured the equivalent width
of \civ $\lambda$1548.20 by directly integrating over the $\pm 25$ \kms
region centered on the position of \civ $\lambda$1548.20, where 
$\pm 25$ \kms corresponds to 3 times the (assumed) FWHM of the
\civ absorption for $b=10$ \kms
based on profile fitting to \civ absorption systems
detected in other HIRES spectra (cf. Rauch et al 1996). We had to assume a
$b$ value for the \civ absorption since no actual detection was
made in any of the composite spectra (section 3.3).

   Because no \civ absorption was detected in the composite spectra,
we estimated the equivalent width limit from Monte Carlo simulations.
For each quasar spectrum, we first picked redshifts 
randomly within the redshift range over which  the actual \lya lines
that were used to form a composite spectrum were found. 
The number of redshifts picked was the same as the observed
number of \lya lines.
We then treated these random redshifts as positions of
\lya lines, assigned a flag to each redshift according to the ``cleanness''
of the \civ region. We then formed a composite spectrum using these
random redshifts and measured the equivalent width at the line
position of \civ $\lambda$1548.20 in exactly the same way as was done
for the real data. This procedure was repeated 4000 times. The
resulting distribution of measured equivalent width at the position
of \civ $\lambda$1548.20 was then examined to infer the 95.5\% confidence
limit on the equivalent width of \civ $\lambda$1548.20 absorption
in the composite spectrum, which corresponds to a 2$\sigma$ upper
limit if the distribution is Gaussian. 
The actual distributions of the measurement equivalent widths from the
simulations closely resemble a Gaussian function.

\subsection{Abundance Estimates}

The 95.5\% confidence limit obtained above for each sample
combined with the expected
mean \nhi\ of the sample (column 5 of Table 4) yields a \nciv/\nhi\
upper limit for the sample.
    In order to turn the N(\textsc{C~iv})/N(\textsc{H~i}) limit
into an estimate of the C abundance in the clouds, one must know the
ionization conditions in the gas. Since only \nhi\ is known,
it is not possible to constrain the ionization
conditions from the ionic ratios. 

    Recent hydrodynamical simulations of cosmological structure
formation in the context of the Cold Dark Matter or other models
have been generally successful in matching the observed characteristics
of \lya forest clouds over the redshift range $2<z<4$
(Cen et al 1994; Petitjean, Mucket, \& Kates 1995;
Zhang, Anninos, \& Norman 1995; Hernquist et al 1995; Miralda-Escude et al
1996; Rauch et al 1997). In these models,
the higher \nhi\ clouds (\nhi$>10^{14}$ \cm2 or so) tend to be
associated with filaments surrounding and connecting collapsed objects, while
the lower \nhi\ clouds tend to be found in voids. 
Rauch et al (1997) addressed the ionization state of the
\lya clouds and heavy elements and showed that the observed 
N(\textsc{C~iv})/N(\textsc{H~i}) vs \nhi\ follows the predicted
relation from the simulations if a mean [C/H]=$-2.5$ is assumed
for the gas. 
Given the lack of observational constraints on the ionization 
conditions in the \lya clouds in our sample and considering the reasonable
success of the simulations, it seems
appropriate to  use their theoretical prediction of 
N(\textsc{C~iv})/N(\textsc{H~i}) vs \nhi\ to
estimate [C/H] for our \lya clouds. An added benefit of using
this relation is that the metallicity of the lower column density
clouds considered here and that of the higher column density
(\nhi$>10^{14.5}$ \cm2) clouds considered by Rauch et al
are derived in a self-consistent manner.

     The N(\textsc{C~iv})/N(\textsc{H~i}) vs \nhi\ relation of
Rauch et al (1997) over $10^{13.5}<$\nhi$<10^{14}$ \cm2 can be described
by the function
$${\rm log N(C IV)/N(H I)}=0.90\times {\rm log N(H I)}-14.80 \eqno(1) $$
for [C/H]$=-2.5$. This relation was used to infer [C/H]
for the various composite spectra discussed in section 3.3.
The conversion from \nciv/\nhi\ to [C/H] was probably the most uncertain step 
in the entire analysis. Unfortunately, it was not possible to
estimate the magnitude
of this uncertainty. However, because the same \nciv/\nhi\
vs \nhi\ relation (i.e., equation (1)) was used to derive metallicities
for both the \nrange clouds (this study) and the \nhi$>10^{14.5}$ \cm2
clouds (Rauch et al), the relative metallicities between these
clouds should be reasonably accurate.

\subsection{Results}

    The results from analyzing the composite spectra for several
samples (next paragraph) are tabulated in Table 4. 
Column 1 of Table 4 gives a letter 
identifying each sample. The second and third columns
give the total number of lines in the sample and the mean redshift
of the lines in the sample. Column 4 indicates the estimated S/N per 6.6 \kms
resolution element in the composite spectrum
based on propagation of the individual error spectra. Column 5
is the expected mean \nhi\ of the sample from the $f(N)\propto N^{-1.5}$
distribution. Column 6 gives the upper limit of the rest frame
equivalent width of \civ $\lambda$1548.20 from the composite spectrum
at the 95.5\% confidence level based on the Monte Carlo simulations
discussed in section 3.1. Column 7 gives the 95.5\% confidence
limit on N(\textsc{C~iv}) derived from the equivalent width limits
in column 6 assuming that such weak lines are on the
linear part of the curve of growth. The last
column gives the metallicity limit at the 95.5\% confidence level
based on the N(\textsc{C~iv})/\nhi\ limit as described in section 3.2.

   Following convention, we treat the clouds within 3,000 \kms of the
emission redshift of the background quasars separately
from those further away from the emission redshift of the quasars
for several reasons: (1) clouds very close to the quasar emission
redshifts may be physically close to the background quasars and may
be more highly ionized due to the enhanced UV ionizing flux.
(2) the heavy element absorption systems within 3,000 \kms of $z_{em}$
often show \textsc{N~v} and relatively high metallicities, suggesting
that many of them may be ejected
from the background quasars or related to their host galaxies or
their immediate environments.

We first formed a composite spectrum using 233 \lya lines in Table 2
that are more than 3,000 \kms way from $z_{em}$ and that
have flag=0; the latter ensures that both the 
\civ $\lambda$1548.20 and $\lambda$1550.77 line regions in the
composite spectrum contain 
contributions from the same \lya systems. No significant \civ 
absorption was seen in the composite spectrum. In order to maximize the
S/N of the composite spectra in the \civ $\lambda$1548.20 region, 
we decided to also include \lya lines with flag=2 in the analysis.

   We considered the following 7 samples:

{\it Sample A}: All \lya lines in Table 2 with flag=0 or 2 that are
more than 3,000 \kms away from $z_{em}$.

{\it Sample B}: All \lya lines in Table 2 with flag=0 or 2 that
are more than 3,000 \kms away from $z_{em}$ and that
have estimated \nhi$<13.75$ (i.e., $10^{13.5}<$\nhi$<10^{13.75}$ \cm2)
 based on the residual flux 
at the line center assuming Doppler $b=25$ \kms.

{\it Sample C}: All \lya lines in Table 2 with flag=0 or 2 that
are more than 3,000 \kms away from $z_{em}$ and that
have estimated \nhi$>13.75$ (i.e., $10^{13.75}<$\nhi$<10^{14}$ \cm2) based
on the residual flux at the line center assuming $b=25$ \kms.

{\it Sample D}: All \lya lines in Table 2 with flag=0 or 2 that 
are more than 3,000 \kms away from $z_{em}$ and that have $z<2.75$.

{\it Sample E}: All \lya lines in Table 2 with flag=0 or 2 that 
are more than 3,000 \kms away from $z_{em}$ and that have $z>2.75$.

{\it Sample F}: All \lya lines in Table 2 with flag=0 or 2
that are more than 3,000 \kms away from $z_{em}$ and that
are within 200 \kms of known metal systems. 

{\it Sample G}: All \lya lines in Table 2 with flag=0 or 2 that are
within 3000 \kms of $z_{em}$.  There are only 12 lines in this sample.

None of the composite spectra showed significant \civ absorption.
The composite spectrum for sample A,
which is the average of 282 \lya lines and which
has an estimated S/N of 1860:1 per 6.6 \kms (the resolution of the
original HIRES spectra), is shown in Figure 3 as an example. 
This S/N is $\sim$10 times better than the best
S/N in the \civ region of individual spectra, and is $\sim 20$ times
better than the {\it average} S/N of the spectra. 
The composite spectrum has a mean flux
extremely close to unity (as expected) but with some mild undulation,
probably as a result of imperfect continuum placement in the original
spectra.

\section{DISCUSSION}

\subsection{Implications of the Lack of Detection of Metals in the Low
Column Density Clouds}

The results tabulated in Table 4 can be summarized briefly as the follows:

(1) Excluding the few \lya clouds that show \civ absorption individually,
\lya clouds with \nrange and $2.2<z<3.6$ have {\it mean} [C/H]$<-3.5$
at the 95.5\% confidence level. This upper limit is a factor of 10 smaller
than the mean [C/H] found for similar \lya clouds with \nhi$>10^{14.5}$ \cm2
(Cowie et al 1995; Tytler et al 1995; Songaila \& Cowie 1996; Rauch et al
1997; Hellsten  et al 1997). The above conclusion appears to hold, 
to within a factor of $\sim 2$,
independent of redshift or \nhi\ for the ranges considered above.

(2) Again, excluding the few clouds that show \civ absorption individually,
\lya clouds with \nrange and $2.2<z<3.6$ have {\it mean} [C/H]$<-3.0$ 
at the 95.5\% confidence level even when they are within 200 \kms of 
known metal systems.

(3) Once again, 
excluding the few clouds that show \civ absorption individually,
\lya clouds with \nrange and $2.2<z<3.6$ have {\it mean} [C/H]$<-2.7$ 
at the 95.5\% confidence level even when they are within 3000 \kms of the
emission redshift of the background quasars. This upper limit is a factor
of $>7-500$ lower than for similar clouds that show \civ absorption
individually (see ``associated'' systems in Table 3 and section 4.2).
It is quite possible that these clouds occur within 3000 \kms of the
emission redshift of the background quasars solely by accident and that
they are not related in anyway to the background quasars. In other words,
they belong to the class of ``intervening'' clouds despite their proximity
in redshift to the background quasars.

The mean metallicity of the \lya clouds with \nrange, [C/H]$<-3.5$, is in
stark contrast to that of the \lya clouds with \nhi$>10^{14.5}$ \cm2
for which several studies have found that 50-75\% of these clouds show 
detectable \civ absorption in individual HIRES spectra with a mean
[C/H]$\simeq -2.5$ (Cowie et al 1995; Tytler et al 1995; Songaila \& Cowie
1996; Rauch et al 1997; Hellsten et al 1997). Surprisingly, 
the \lya clouds with 
$10^{13.75}<$\nhi$<10^{14}$ \cm2, with a mean \nhi\ that is only a factor
of $\sim 4$ smaller than the limiting \nhi$=10^{14.5}$ \cm2 probed by 
previous studies, are found to have a mean [C/H]$<-3.55$ or a factor of
11 lower. These results suggest that a rapid decrease with decreasing
\nhi\ in the mean metallicity level of the clouds sets in at
\nhi$\sim 10^{14}-10^{14.5}$ \cm2.

In figure 4 we show very crudely the metallicity distribution of quasar
absorption line systems as a function of \nhi. The effects of non-solar
relative abundances are ignored. A trend of decreasing 
metallicity with decreasing \nhi\ is indicated by figure 4. 
Since the higher column
density systems are likely to be more intimately associated with collapsed 
objects, where star formation is expected to occur earlier and more
vigorously, the above trend of decreasing metallicity with 
decreasing \nhi\ is consistent with the scenario in which star formation
in collapsed objects eject metal-enriched material outwards and pollute
their surroundings.

The detection of \civ absorption in individual \lya clouds with \nhi\ as 
low as $10^{14.5}$ \cm2 in high sensitivity Keck observations was
considered evidence for the occurrence of an early generation of stars
(Pop III) that may have polluted the entire universe in a wide spread
fashion to a nearly uniform level of [C/H]$\simeq -2.5$ 
(Songaila \& Cowie 1996).
A similar inference was made by Lu, Sargent, \& Barlow (1997) based on
the rough agreement between the metallicity, [C/H]$\simeq -2.5$, that was 
inferred for the \lya clouds and that of the lowest metallicity damped
\lya galaxies, generally believed to be the high-redshift progenitors of
normal galaxies (Wolfe 1988). However, the extremely low metallicities we
infer for the \lya clouds with \nrange rule out the scenario of uniform
contamination by Pop III stars to [C/H]$\sim -2.5$. Recent theoretical
studies appear to corroborate this results. For example,
Ostriker \& Gnedin (1996) and Gnedin \& Ostriker (1997;
see also Murakami \& Yamashita 1997; Steinmetz 1997)
recently explored the process of early metal enrichment
using high resolution numerical
simulations. In the COBE-normalized Cold Dark Matter plus cosmological 
constant model investigated by these authors, the first generation of
stars (Pop III) occurred at a redshift of $z\sim 14$ through molecular
hydrogen cooling. Subsequent supernova explosions from such stars then
enriched the universe to a mean metallicity of $10^{-3.7}$ solar, which
is considerably lower than the mean [C/H]$\simeq -2.5$ found for
\lya\ clouds with \nhi$>10^{14.5}$ \cm2 but is consistent with the upper
limit found for \lya clouds with \nrange.
The energy release from the supernova also raised the temperature
of the gas and destroyed
the hydrogen molecules through photodissociation, thus temporarily shutting
down the star formation process. It was not until $z<10$ when hydrogen
line cooling became sufficient to induce a second major episode of star 
formation (Pop II). Interestingly, Gnedin \& Ostriker predicted that
there should be a sharp drop in the metallicity level of \lya\ clouds 
at column density below about $10^{13.5}-10^{14.5}$ \cm2 at $z=4$ 
(see their figure 7), a feature apparently confirmed by our result.

In view of all the evidence, we conclude that, if a generation 
of Pop III stars
occurred in the very early universe,  the mean enrichment level of
the intergalactic medium resulting from the associated supernova should
be [C/H]$<10^{-3.5}$. The relatively high metallicities, [C/H]$\simeq -2.5$,
found for the \nhi$>10^{14.5}$ \cm2 clouds are best explained by
later enrichment from Pop II stars occurring either within the clouds 
themselves or in nearby galaxies.

It is possible to obtain some clues about the size scale  of the
metal-enriched (i.e., [C/H]$\simeq -2.5$) regions.
If we assume that the observed velocity spread $\Delta v$
of the components in \civ systems are purely due to ejection of 
metal-enriched gas from supernova, we can set a conservative upper limit
on the maximum distance this enriched gas could have traveled by
the time of observations. This
distance is of the order of 200 kpc if a typical $\Delta v=400$ \kms
is adopted and if it is assumed that the first supernova occurred
$\sim 1$ Gyrs before our observation ($z\sim 3$).
The lack of detectable metals in \lya\ clouds
within 200 \kms of known metal systems puts another limit on the spatial
size of the enriched region to be $<500 h^{-1}_{50}$ kpcs if it is
assumed that this velocity difference is purely due to Hubble flow.
Finally, according to simulations, the \nhi$=10^{14}$ \cm2 contour,
which roughly separates metal-rich gas with [C/H]$\simeq -2.5$ from
metal-poor gas with [C/H]$<-3.5$, delineates a
continuous filamentary gaseous structure in which galaxies are embedded.
The thickness of this structure is of the order
100 kpcs (Rauch et al 1997), in broad agreement with other estimates.

\subsection{Metallicity and Nature of Low N(HI) \lya 
Systems with Detectable C IV Absorption}

As noted briefly in section 2.2, some of the low column density \lya clouds
considered here
show detectable \civ absorption individually (figure 2 and Table 3). 
This is in stark contrast with the vast majority of systems 
with similar \nhi\ whose composite spectra did
not reveal any \civ absorption down to very sensitive limits (Table 4).

     To understand the nature of these low \nhi\ \lya clouds that apparently
contain a significant amount of metals, we estimated the lower limit to
their metallicity adopting the usual assumption that the clouds are
photoionized by the integrated light from quasars. We ran several set
of photoionization models using CLOUDY v90.02 (Ferland 1996) assuming
plane-parallel geometry and uniform gas density for the clouds. We
considered several plausible ionizing spectra: (1) the mean quasar spectrum
constructed by Haardt \& Madau (1996; hereafter HM) assuming a transparent
universe; the spectrum has a spectral index
$\alpha=-1.5$ over the critical 1-4 Rydberg region;
(2) the above spectrum after taking into account the opacity of intervening
clouds as calculated by HM (their spectrum at redshift $z=3$ is adopted);
Two further ionizing spectra considered were similar to the two 
mentioned above
but with $\alpha=-1.8$ over 1-4 Rydberg region.
These latter spectra, which were also from HM (private communication), 
were calculated 
in light of the work by Zheng et al (1998) and Laor et al (1997) 
after the HM paper had appeared in press. 
The CLOUDY results gave predictions about
the variation of \nciv\ as a function of the ionization parameter,
$\Gamma$ (the ratio of ionizing photon density to 
hydrogen gas particle density)
for chosen \nhi\ and metallicity. 
For a given system with measured \nhi\ and \nciv,
we estimated a lower limit to [C/H] by assuming that the 
\textsc{C~iv}/\textsc{H~i} ratio is at the peak of the distribution.
This approach resulted in the [C/H] lower limits  given
in column 7 ([C/H]$_{\rm P}$) of Table 3, 
where the CLOUDY results based on the $\alpha=-1.8$
HM spectrum without intervening opacity was used for the ``associated''
systems, and the CLOUDY results based on the $\alpha=-1.8$ HM spectrum
taking into account the intervening opacity was used for the ``intervening''
systems. The ratio of \textsc{C~iv}/\textsc{H~i} typically peaks
at $\Gamma=-1\pm0.5$ in these models.  The [C/H]
lower limits were $\sim 0.4$ dex higher than these given in Table 3 for the
``intervening'' systems if the $\alpha=-1.5$ HM spectrum with intervening
opacity was adopted. Similarly, the [C/H] lower limits would be
$\sim 0.2$ dex higher for the ``associated'' systems if the $\alpha=-1.5$
HM spectrum without intervening opacity was adopted. Accordingly,
the [C/H]$_{\rm P}$ limits given in Table 3 are relatively 
conservative estimates.\footnote{Incidently, if we use the relation between
log \nciv/\nhi\ and log \nhi\ from Rauch et al  (1997; equation (1) of
this paper) to estimate the
actual metallicity of the intervening systems, as was done for the
composite spectra, the resulting [C/H]$_{\rm P}$ values (no longer
lower limits) will be 0.05
to 0.35 dex higher than the lower limits quoted in Table 3 for these
systems. The relative small differences suggest that the clouds in
the Rauch et al simulations are near the peak of \textsc{C~iv}/\textsc{H~i} 
distribution.}

   It is possible that the clouds given in Table 3 
are collisionally ionized rather than photoionized. 
Earlier studies (Blades 1988, and references therein; Rauch et al 1996)
found that the line widths of many \civ\ absorption
lines are too narrow to be collisionally ionized. However, the conditions
in the very low column density clouds considered here could be different .
For gas that is in collisional ionization equilibrium, the
ratio of \textsc{C~iv}/\textsc{H~i} peaks at temperature $T=10^{5.05}$ K
with \textsc{C~iv}/\textsc{H~i}=$10^{4.3}$C/H (Sutherland \& Dopita 1993).
For solar metallicity, this implies \textsc{C~iv}/\textsc{H~i}=7.2.
In comparison, the peak \textsc{C~iv}/\textsc{H~i} value is around
unity for the photoionization models considered above when solar 
metallicity is assumed. Thus collisionally
ionized clouds could reach much higher \textsc{C~iv}/\textsc{H~i}  ratio
than photoionized clouds for the same metallicity. The [C/H] limits
derived from collisional ionization assuming the 
peak \textsc{C~iv}/\textsc{H~i} ratio are given
in column 8 ([C/H]$_{\rm C}$) of Table 3. However, the widths of all but
one of the \lya absorption lines imply $T<10^{4.7}$ K, which is
inconsistent with these clouds being collisionally ionized
at $T\sim 10^{5}$ K. It could be argued
that the observed \lya absorption is not responsible for the \civ
absorption. In this case, the \lya absorption corresponding to the 
collisionally-ionized \civ absorption
could be broad ($b=41$ \kms for $T=10^{5}$ K) and shallow and may be
hidden for some of the systems.  However, 3 of the 7 
intervening systems in Table 3 have temperature $T<10^{4.6}$ K based
on the width of the \civ absorption lines. Similarly, 2 of 8 of the
associated systems in Table 3 have $T<10^{4.6}$ K based on the
width of the \civ absorption. Clearly the \civ in these clouds
cannot be produced by collisional ionization at $T\sim 10^5$ K
under equilibrium conditions. 
It is perhaps no coincidence that almost all the \civ absorption lines
with inferred $T<10^{5}$ K are relatively strong and well measured, while
those with inferred $T>10^{5}$ K are all relatively weak, suggesting
that the temperature of the latter systems may have been
overestimated owing to our inability to discern the component structure.
We therefore conclude the these clouds are not likely to be collisionally
ionized under equilibrium conditions.

However, the clouds which show detectable \civ\ absorption
could be collisionally ionized under non-equilibrium
conditions. Indeed, hot plasma gas cools very fast over the $T\sim 10^5$ K
regime. As shown in figure 15 of Sutherland \& Dopita (1993), the cooling
time for gas with solar composition at $T\sim 10^5$ K is shorter 
than than the time to reach collisional ionization 
equilibrium for \civ and especially for \textsc{C~v}. 
Consequently, the overall ionization balance of C is governed
by the slowly recombining \textsc{C~v} ion. Figure 16 of Sutherland \& Dopita
indicates that large fractions of \civ, \textsc{C~v}, and \textsc{C~vi} ions
can exist at temperature as low as $10^{4.5}$ K under non-equilibrium
conditions, while under collisional ionization equilibrium the fractions
of these ions at such a low temperature would be negligible. 
These considerations
raise the possibility that the \civ ions in the \lya clouds listed in Table 3
with $T<10^{4.6}$ K could be produced by collisional ionization but under
non-equilibrium conditions. However, figure 16 of Sutherland \& Dopita
also indicates that N(\textsc{C~ii})$>$\nciv\ under such conditions. When the
likely sub-solar metallicities of the \lya clouds are taken into account, we
infer from figures 9, 14, and 15 of Sutherland \& Dopita that the actual
cooling time scale in the clouds is likely to be shorter than the
collisional ionization equilibrium time scale of \textsc{C~ii} and
\textsc{C~v} but longer than that of \civ. This should further {\it increase}
N(\textsc{C~ii}) relative to N(\textsc{C~iv}) because more \civ should 
recombine to \textsc{C~iii} than can be created through the recombination
of \textsc{C~v}. As a result, we expect to detect \textsc{C~ii} $\lambda$1334
absorption in these \lya systems if the \civ ions were
produced by collisional ionization at $T<10^{4.6}$ K under non-equilibrium
conditions. No \textsc{C~ii} $\lambda$1334 absorption is found in any of
the systems listed in Table 3
in our HIRES spectra (the \textsc{C~ii} $\lambda$1334 regions
for three of the systems are contaminated by unrelated absorption lines).

   It therefore seems likely that the clouds showing detectable \civ
absorption are photoionized
rather than collisionally ionized. Assuming this is the case, one
then find that the intervening clouds have metallicities ranging
from [C/H]$>-2.2$ to [C/H]$>-0.9$ (see column 7 of Table 3), 
which is significantly higher
than for typical \lya clouds of similar \nhi. 
Several of the associated systems have [C/H] within
a factor of 2 of the solar metallicity.
This result is consistent with previous work (Savaglio, D'Odorico, \& Moller
1994; Petitjean, Rauch, \& Carswell 1994; Tripp, Lu, \& Savage 1996)
and it  indicates rapid enrichment in the
quasar nuclei/environment. The same characteristics appear to
be shared by the broad absorption line clouds and broad emission
line clouds (cf. Hamann 1997, and references therein).
A few of the what we classify as ``associated'' systems
(because they are within 3,000 \kms of $z_{em}$) actually have [C/H]
lower limits comparable to most of the intervening systems; some of these
could be clouds that are unrelated to the quasar environment but happen
to have redshifts close to $z_{em}$. 

The inferred lower limits of [C/H] for the intervening clouds listed in
Table 3 are factors 20-400 (or more)  higher than the mean [C/H]$<-3.5$
 found for the vast majority of intervening clouds (Table 4).
What are the causes of such large differences?
We consider two possibilities: 
(1) the intervening clouds which show individual \civ absorption
could represent the high metallicity tail of the metallicity 
distribution of the \lya clouds; 
(2) the intervening clouds with detectable \civ absorption
have a separate origin:  perhaps they are ejected from the
background quasar despite the large ejection velocities 
($>10,000$ \kms) required, or they may be ejected from foreground quasars
near the quasar sight line, in which case extremely large ejection
velocity are no longer required. The latter possibility
of case (2) above
could be check by searching  for foreground quasars near the redshifts
of the \lya systems that show detectable \civ absorption.
However, that the ``intervening'' systems listed in Table 3 seem to have
lower \textsc{C~iv}/\textsc{H~i} ratios  on average than the ``associated''
systems suggests that they are probably not ejected from quasars.

    We now consider in some detail the more interesting case (1) above.
If the weak \lya lines in our sample have a broad distribution in
metallicity and if the intervening clouds showing \civ
absorption indeed represent the high metallicity tail of that distribution,
two conditions should be met.  Take sample A for example.
First, the metallicity distribution in the clouds must be such
that 6\footnote{One of the intervening systems listed in Table 3 ($z=3.13247$
toward Q1422+2309) is within 50 \kms of a strong \lya line and does not
qualify to be included in our sample.} 
of 282+6 randomly-drawn systems should have [C/H]$>-2.17$.
Second, the composite spectrum of the remaining 282 systems must not
show \civ $\lambda$1548 absorption with $w_r(1548)>0.088$ m\AA\
or \nciv$>10^{10.34}$ \cm2 (Table 4).

   In order to examine the above conditions quantitatively, we
explore the following simple model. We
assume that the [C/H] distribution of the clouds can be
described by a Gaussian function with a mean metallicity $M_{0}$
and a standard dispersion $\sigma_M$. The first condition, that
6 of 288 randomly-drawn systems have [C/H]$>-2.17$ 
then requires that [C/H]$=-2.17$
is 2.04$\sigma_M$ away from $M_0$. In other words:
$$-2.17-M_0=2.04\sigma_M \eqno(2).$$
For the second condition, it can be shown that, for a ensemble of
clouds drawn randomly from the H I column density distribution
$f(N)\propto N^{-1.5}$ and the above metallicity distribution,
the mean column density of \civ is
$${\rm \nciv}=\sum {\rm N(HI)}^{1.9}\times 10^{{\rm [C/H]}-12.30}/N_l,
 \eqno(3)$$
where $N_l$ is the total number of lines in the sample. Here, equation
(1) has been used to derive \nciv\ for individual cloud from
the \nhi\ and [C/H] values drawn from the distributions. The sum in
equation (3) can be estimated numerically. 
 Strictly speaking, one should take into account the variable detection
sensitivity (i.e., variable S/N) of the spectra in the above analysis.
However, given the crudeness of this analysis, and
considering the relative uniformity of the HIRES spectra, assuming
a uniform detection sensitivity for this analysis probably suffice for
our purposes.

    Equation (2) maps out a straight line in the $M_0$ vs $\sigma_M$
plane, while equation (3) combined with the requirement
\nciv$<10^{10.34}$ \cm2 maps out a region in the  $M_0$ vs $\sigma_M$
plane. The results are shown in Figure 5, where the solid straight line
is equation (2). The shaded area indicates the [$M_0$, $\sigma_M$]
combinations for which the composite spectrum of 282 randomly drawn
\lya systems, subject to the condition [C/H]$<-2.17$, would not reveal
a \civ $\lambda$1548.20 absorption with mean \nhi$>10^{10.34}$ \cm2
at the 90\% confidence level. The solid line
connecting the dots gives the  [$M_0$, $\sigma_M$] relation for which the
composite spectrum of 282 randomly drawn \lya systems (again, subject
to the condition [C/H]$<-2.17$) would produce a \civ $\lambda$1548.20
absorption with a mean \nciv$=10^{10.34}$ \cm2. It is seen that only at
$\sigma_M>1$ or so is there any overlap between the straight line and
the shaded area, when the two conditions mentioned above are met
simultaneously. Thus, within the validity of the simple model explored
here, if the 7 intervening \lya systems with \nrange that individually
show \civ absorption are drawn from the same population as the remaining
\lya systems with similar \nhi\ which do not show detectable \civ\
absorption, then the overall
metallicity distribution of the clouds must be such
that it has a mean metallicity [C/H]$<-4$ with a dispersion $>1$ dex.
This is roughly consistent with the mean metallicity of $10^{-3.7}$
predicted by Ostriker \& Gnedin (1996) for the enrichment by
Pop III stars.

\section{SUMMARY AND CONCLUSIONS}

   We have investigated the metal contents of \lya clouds at $2.2<z<3.6$ with 
\nrange\ using high resolution (FWHM=6.6 \kms), high S/N ($\sim$100:1)
spectra obtained with the
10m Keck I telescope. Previous investigations of similar nature were
limited to \lya clouds with \nhi$>10^{14.5}$ \cm2 for which a mean
metallicity of [C/H]$\simeq -2.5$ was deduced. It has been suggested
that the metals seen in these clouds may have been produced by a generation 
of Population III stars occurred at a much earlier epoch before the 
formation of quasars and normal galaxies. 
From 9 of our best S/N quasar spectra obtained with the HIRES,
we selected $\sim$300 \lya lines with estimated \nhi\ between \nrange.
The \lya lines selected
have absorption redshift $2.2<z_{abs}<3.6$ (excluding \lya lines within
3000 \kms of the quasar emission redshifts).
After eliminating a small number
of \lya lines that individually show detectable \civ absorption (see point 4
below), we shifted the quasar spectra into the rest frame
of each of the remaining \lya clouds in the sample and then averaged
the rest-frame spectra over the spectral region encompassing the \civ
$\lambda\lambda$1548.20, 1550.77 absorption lines.
We reached S/N as high as 1860:1 per
6.6 \kms resolution element in our composite spectra. Such a high S/N
would have been impossible to obtain through direct observations.
The main conclusions are as follows:

(1) No significant \civ absorption was
detected in any of the composite spectra we investigated. The most 
sensitive limit was obtained when all \lya lines in the sample (excluding
those showing \civ absorption individually) that are more than 3,000 \kms
away from the background quasars were
included in the analysis, for which we reached a S/N=1860:1 per
resolution element, corresponding to  an upper limit of $<0.088$
m\AA\ (95.5\% confidence limit) for the rest-frame equivalent width of the
\civ $\lambda$1548.20 line. We deduced a mean metallicity of [C/H]$<-3.5$
for these clouds using the cosmological simulations of Rauch et al (1997) to 
infer the ionization correction; the same simulation results have
been used previously to deduce a mean metallicity 
of [C/H]$\simeq -2.5$ for \lya clouds
with \nhi$>10^{14.5}$ \cm2. The low metallicities derived for the
\nrange clouds appear to hold to within a factor of $\sim 2$ independent
of redshift or \nhi\ over the range considered above.
In particular, a sample of \lya clouds occurring within 200 \kms of known
metal systems yielded a mean metallicity of [C/H]$<-3.0$ (95.5\% confidence
limit) based on a composite spectrum analysis.

(2) The mean metallicity of \lya clouds with \nrange, [C/H]$<-3.5$, 
is a factor of 10 lower than that inferred for \lya clouds with
\nhi$>10^{14.5}$ \cm2. This suggests that a sharp drop in the metallicity
level of \lya\ clouds sets in at \nhi$=10^{14}-10^{14.5}$ \cm2.
This result rules out the suggestion that a generation of Pop III stars
could have polluted the entire universe to a uniform level of 
[C/H]$\simeq -2.5$.
Cosmological simulations involving  gas hydrodynamics indicate that
\lya absorption 
with \nhi$>10^{14.5}$ \cm2 mostly occurs in continuous filaments
of gas surrounding and connecting  collapsed objects, while those
with \nhi$<10^{14}$ \cm2 are preferentially found in void regions further 
away from collapsed objects. These results, coupled with theoretical
predictions about Pop III star formation and enrichment (Ostriker \& Gnedin
1996; Gnedin \& Ostriker 1997), strongly suggest that most of the
heavy elements in \lya clouds with \nhi$>10^{14.5}$ \cm2 were probably
produced {\it in situ} by Pop II stars, 
in the sense that they were either made by stars within
the clouds or were ejected from nearby star-forming galaxies.
Consequently, the void regions (\nhi$<10^{14}$ \cm2) could only
have experienced the enrichment by Pop III stars.
We estimated a typical thickness of the order of 100 kpc
at $z=3$ for the metal-enriched filamentary structures.

(3) The low metallicities we inferred for 
the \lya clouds with \nrange require that,
if there was a generation of Pop III stars that occurred
in the very early universe, 
the mean enrichment level resulting from the explosions of such stars should
be [C/H]$<-3.5$ for \lya clouds with \nrange at $2.2<z<3.6$, or 
[C/H]$<-4$ if the \lya clouds which individually show detectable 
\civ absorption are to be considered part of the overall metallicity 
distribution (point 4 below). These results are consistent with the
recent theoretical calculations of Ostriker \& Gnedin (1996).

(4) A small fraction (16) of the \lya clouds with \nrange show
detectable \civ absorption individually (figure 2 and Table 3). 
Of the 16 systems, 9 occur within
3,000 \kms of the emission redshift of the background quasars and may
be gas ejected from the quasars or gas associated with the quasar host
galaxies and/or their environment. Most of these clouds are inferred to have
[C/H] close to or even exceeding the solar value based on photoionization
models, consistent with previous studies. The remaining 7
systems have velocities $>10,000$ \kms away from the emission redshift
of the background quasars and are likely to be intervening in nature.
Collisional ionization is ruled out for at least 3 of the 7 intervening
clouds based on the width of the \civ absorption lines. If the clouds
are photoionized by the integrated light from quasars, as is commonly
assumed, the lower limits to their [C/H] values are inferred to be
between $-2.2$ and $-0.9$. The relatively high metallicities of these
clouds are in stark contrast with those of other similar \nhi\ clouds that 
do not show \civ absorption individually for which we inferred a mean
[C/H]$<-3.5$ from their composite spectrum. If \lya clouds with \nrange have 
a  metallicity distribution that can be approximated by a Gaussian, and if
these clouds that individually show detectable \civ absorption represent
the high metallicity tail of this distribution,
we infer that the metallicity
distribution should have a mean [C/H]$<-4$ with a dispersion of at least
1 dex. Alternatively, the \lya clouds which individually show detectable
\civ absorption may have a separate origin.

\acknowledgements
We thank Francesco Haardt and Piero Madau for providing their quasar ionizing
spectra in electronic form, and Gary Ferland for a copy of his CLOUDY code.
We also thank all of the many people who made the Keck telescopes and their
instrumentation possible.
WLWS acknowledges support from NSF grant AST95-29073.
MR acknowledges support from NASA through grant number 
HF1075.01-94A from the
Space Telescope Science Institute, which is operated by the Association
of Universities for Research in Astronomy, Inc., for NASA under contract
NAS5-26555. 

\begin{planotable}{crrrl}
\tablewidth{40pc}
\tablecaption{List of Quasars}
\tablehead{
\colhead{QSO}              &\colhead{$z_{em}^a$}    
 &\colhead{V$^b$}          &\colhead{S/N$^c$}
 &\colhead{$\lambda$ Range$^d$}
 }
\startdata
0100$+$1300 &2.718 &16.6 &80-110 &4090-4520  \nl
0636$+$6801 &3.179 &16.5 &80-120 &4286-5080  \nl
1107$+$4847 &2.965 &16.7 &60-90 &4235-4820   \nl
1422$+$2309 &3.629 &16.5 &90-160 &4748-5627  \nl
1425$+$6039 &3.175 &16.5 &90-160 &4282-5075$^e$  \nl
1442$+$2931 &2.661 &16.2 &70-100 &3900-4450$^e$  \nl
1700$+$6416 &2.744 &16.1 &80-120 &4000-4551      \nl
1946$+$7658 &3.053 &15.9 &80-130 &4157-4927$^e$  \nl
2343$+$1232 &2.579 &17.0 &80-110 &3940-4351$^e$  \nl
\enddata

\tablenotetext{a}{Adopted emission redshift for the quasar. The emission
redshift is determined either from the peak of the \lya emission
line in the HIRES spectrum or from the highest-redshift \lya
absorption line observed in the spectrum, whichever is larger.}

\tablenotetext{b}{Visual magnitude of the quasar.}

\tablenotetext{c}{Typical S/N per  6.6 \kms
resolution element in the \civ region.}

\tablenotetext{d}{Wavelength region used to select \lya lines (see
text). The lower limit is either determined by S/N requirement
or by the onset of the  Ly$\beta$ emission line of the quasar. 
The upper limit is determined by the \lya emission line.}

\tablenotetext{e}{The following spectral regions are excluded 
owing to damped Ly$\alpha$ absorption: 4643-4663 (Q 1425+6039),
4168-4188 (Q 1442+2931), 4660-4683 (Q1946+7658),
4155-4185 (Q 2343+1232).}

\end{planotable}

\begin{planotable}{ccccc}
\tablewidth{40pc}
\tablecaption{List of Ly$\alpha$ Systems}
\tablehead{
\colhead{Wavelength}       &\colhead{Redshift}    
 &\colhead{Residual Flux}  &\colhead{Flag} }
\startdata
\cutinhead{Q 0100+1300}
  4094.153 &   2.36782 &  0.122 &  1  \nl 
  4100.934 &   2.37339 &  0.180 &  0  \nl 
  4116.390 &   2.38611 &  0.131 &  0  \nl 
  4119.285 &   2.38849 &  0.335 &  0  \nl 
  4122.550 &   2.39118 &  0.054 &  0  \nl 
  4127.273 &   2.39506 &  0.039 &  0  \nl 
  4130.397 &   2.39763 &  0.238 &  0  \nl 
  4132.794 &   2.39960 &  0.274 &  0  \nl 
  4133.622 &   2.40028 &  0.353 &  2  \nl 
  4143.128 &   2.40810 &  0.115 &  0  \nl 
  4174.774 &   2.43413 &  0.373 &  0  \nl 
  4177.658 &   2.43651 &  0.089 &  0  \nl 
  4235.625 &   2.48419 &  0.108 &  0  \nl 
  4258.374 &   2.50290 &  0.275 &  0  \nl 
  4282.304 &   2.52259 &  0.173 &  0  \nl 
  4285.318 &   2.52507 &  0.258 &  0  \nl 
  4291.262 &   2.52996 &  0.234 &  0  \nl 
  4301.424 &   2.53832 &  0.132 &  0  \nl 
  4321.220 &   2.55460 &  0.367 &  2  \nl 
  4354.237 &   2.58176 &  0.348 &  0  \nl 
  4356.169 &   2.58335 &  0.107 &  0  \nl 
  4357.232 &   2.58422 &  0.058 &  0  \nl 
  4359.902 &   2.58642 &  0.319 &  0  \nl 
  4361.825 &   2.58800 &  0.372 &  0  \nl 
  4362.708 &   2.58873 &  0.107 &  0  \nl 
  4380.022 &   2.60297 &  0.122 &  0  \nl 
  4382.571 &   2.60507 &  0.283 &  0  \nl 
  4396.414 &   2.61645 &  0.361 &  0  \nl 
  4422.073 &   2.63756 &  0.401 &  0  \nl 
  4428.246 &   2.64264 &  0.269 &  0  \nl 
\cutinhead{Q 0636+6801}
  4286.281 &   2.52586 &  0.340 &  0  \nl 
  4287.469 &   2.52684 &  0.210 &  0  \nl 
  4289.109 &   2.52818 &  0.298 &  0  \nl 
  4308.544 &   2.54417 &  0.307 &  0  \nl 
  4311.484 &   2.54659 &  0.056 &  0  \nl 
  4331.696 &   2.56322 &  0.098 &  2  \nl 
  4338.049 &   2.56844 &  0.163 &  2  \nl 
  4391.302 &   2.61225 &  0.106 &  0  \nl 
  4421.211 &   2.63685 &  0.308 &  0  \nl 
  4423.604 &   2.63882 &  0.036 &  0  \nl 
  4436.713 &   2.64960 &  0.256 &  0  \nl 
  4440.883 &   2.65303 &  0.204 &  0  \nl 
  4454.959 &   2.66461 &  0.218 &  0  \nl 
  4471.645 &   2.67834 &  0.203 &  0  \nl 
  4482.473 &   2.68724 &  0.065 &  0  \nl 
  4490.741 &   2.69405 &  0.151 &  0  \nl 
  4507.129 &   2.70753 &  0.098 &  1  \nl 
  4536.054 &   2.73132 &  0.853 &  1  \nl 
  4550.071 &   2.74285 &  0.286 &  0  \nl 
  4555.496 &   2.74731 &  0.042 &  0  \nl 
  4577.281 &   2.76523 &  0.036 &  0  \nl 
  4633.240 &   2.81126 &  0.219 &  2  \nl 
  4654.988 &   2.82915 &  0.321 &  0  \nl 
  4668.517 &   2.84028 &  0.124 &  0  \nl 
  4673.233 &   2.84416 &  0.051 &  2  \nl 
  4677.685 &   2.84783 &  0.208 &  0  \nl 
  4682.170 &   2.85151 &  0.335 &  1  \nl 
  4699.186 &   2.86551 &  0.154 &  0  \nl 
  4734.244 &   2.89435 &  0.322 &  1  \nl 
  4735.202 &   2.89514 &  0.339 &  3  \nl 
  4750.998 &   2.90813 &  0.260 &  0  \nl 
  4752.244 &   2.90916 &  0.097 &  1  \nl 
  4765.640 &   2.92018 &  0.091 &  0  \nl 
  4789.042 &   2.93942 &  0.114 &  0  \nl 
  4790.069 &   2.94027 &  0.250 &  0  \nl 
  4799.148 &   2.94774 &  0.144 &  0  \nl 
  4800.392 &   2.94876 &  0.373 &  0  \nl 
  4805.420 &   2.95290 &  0.115 &  0  \nl 
  4827.638 &   2.97117 &  0.094 &  0  \nl 
  4830.792 &   2.97377 &  0.098 &  0  \nl 
  4840.324 &   2.98161 &  0.171 &  0  \nl 
  4849.687 &   2.98931 &  0.342 &  0  \nl 
  4857.483 &   2.99572 &  0.219 &  0  \nl 
  4860.029 &   2.99782 &  0.286 &  0  \nl 
  4890.703 &   3.02305 &  0.046 &  0  \nl 
  4942.479 &   3.06564 &  0.191 &  0  \nl 
  4962.583 &   3.08218 &  0.091 &  2  \nl 
  4968.049 &   3.08668 &  0.085 &  0  \nl 
  4970.790 &   3.08893 &  0.273 &  1  \nl 
  4977.749 &   3.09466 &  0.085 &  0  \nl 
  5005.265 &   3.11729 &  0.115 &  0  \nl 
  5018.773 &   3.12840 &  0.316 &  0  \nl 
  5033.037 &   3.14013 &  0.224 &  0  \nl 
  5034.347 &   3.14121 &  0.243 &  0  \nl 
\cutinhead{Q 1107+4847}
  4239.687 &   2.48753 &  0.090 &  0  \nl 
  4256.232 &   2.50114 &  0.193 &  2  \nl 
  4257.300 &   2.50202 &  0.171 &  0  \nl 
  4299.297 &   2.53657 &  0.357 &  0  \nl 
  4304.639 &   2.54096 &  0.283 &  0  \nl 
  4347.466 &   2.57619 &  0.216 &  0  \nl 
  4364.434 &   2.59015 &  0.060 &  0  \nl 
  4380.748 &   2.60357 &  0.136 &  0  \nl 
  4391.133 &   2.61211 &  0.049 &  0  \nl 
  4398.393 &   2.61808 &  0.067 &  0  \nl 
  4405.924 &   2.62428 &  0.072 &  2  \nl 
  4426.362 &   2.64109 &  0.291 &  1  \nl 
  4433.019 &   2.64656 &  0.103 &  0  \nl 
  4438.663 &   2.65121 &  0.055 &  0  \nl 
  4441.819 &   2.65380 &  0.249 &  0  \nl 
  4466.188 &   2.67385 &  0.142 &  1  \nl 
  4473.093 &   2.67953 &  0.287 &  0  \nl 
  4484.903 &   2.68924 &  0.219 &  0  \nl 
  4496.837 &   2.69906 &  0.310 &  0  \nl 
  4507.029 &   2.70744 &  0.354 &  1  \nl 
  4507.991 &   2.70824 &  0.383 &  2  \nl 
  4508.903 &   2.70899 &  0.133 &  3  \nl 
  4522.404 &   2.72009 &  0.064 &  0  \nl 
  4525.017$^a$ &   2.72224 &  0.083 &  3  \nl 
  4542.543 &   2.73666 &  0.037 &  0  \nl 
  4544.707 &   2.73844 &  0.353 &  0  \nl 
  4556.674 &   2.74828 &  0.372 &  0  \nl 
  4558.005 &   2.74938 &  0.327 &  0  \nl 
  4559.907 &   2.75094 &  0.049 &  0  \nl 
  4568.649 &   2.75813 &  0.100 &  0  \nl 
  4588.396 &   2.77438 &  0.106 &  0  \nl 
  4593.184 &   2.77832 &  0.238 &  0  \nl 
  4628.681 &   2.80751 &  0.053 &  0  \nl 
  4639.846 &   2.81670 &  0.209 &  2  \nl 
  4654.944 &   2.82912 &  0.057 &  0  \nl 
  4668.535 &   2.84030 &  0.302 &  0  \nl 
  4669.555 &   2.84114 &  0.113 &  0  \nl 
  4682.405 &   2.85171 &  0.350 &  0  \nl 
  4696.837 &   2.86358 &  0.249 &  2  \nl 
  4709.816 &   2.87426 &  0.076 &  0  \nl 
  4711.625 &   2.87574 &  0.056 &  0  \nl 
  4739.227 &   2.89845 &  0.368 &  0  \nl 
  4740.343 &   2.89937 &  0.086 &  0  \nl 
  4759.047 &   2.91475 &  0.238 &  1  \nl 
  4783.936 &   2.93523 &  0.194 &  0  \nl 
  4809.364$^a$ &   2.95614 &  0.044 &  3  \nl 
  4813.176 &   2.95928 &  0.301 &  0  \nl 
\cutinhead{Q 1422+2309}
  4758.797 &   2.91455 &  0.237 &  0  \nl 
  4766.250 &   2.92068 &  0.415 &  0  \nl 
  4777.278 &   2.92975 &  0.070 &  1  \nl 
  4790.863 &   2.94092 &  0.085 &  2  \nl 
  4792.363 &   2.94216 &  0.359 &  0  \nl 
  4811.781 &   2.95813 &  0.337 &  0  \nl 
  4819.289 &   2.96431 &  0.099 &  2  \nl 
  4821.130 &   2.96582 &  0.141 &  3  \nl 
  4826.135 &   2.96994 &  0.175 &  2  \nl 
  4840.907 &   2.98209 &  0.270 &  1  \nl 
  4876.051 &   3.01100 &  0.277 &  0  \nl 
  4880.510 &   3.01467 &  0.100 &  2  \nl 
  4896.016 &   3.02742 &  0.290 &  0  \nl 
  4899.455 &   3.03025 &  0.161 &  3  \nl 
  4910.845 &   3.03962 &  0.210 &  2  \nl 
  4917.801 &   3.04534 &  0.330 &  1  \nl 
  4918.696 &   3.04608 &  0.224 &  1  \nl 
  4919.782 &   3.04697 &  0.156 &  0  \nl 
  4921.119 &   3.04807 &  0.235 &  0  \nl 
  4930.662 &   3.05592 &  0.100 &  2  \nl 
  4937.801 &   3.06179 &  0.364 &  0  \nl 
  4948.732 &   3.07079 &  0.278 &  3  \nl 
  4982.914 &   3.09890 &  0.152 &  0  \nl 
  4987.255 &   3.10247 &  0.035 &  2  \nl 
  4991.231 &   3.10575 &  0.252 &  2  \nl 
  4993.338 &   3.10748 &  0.315 &  0  \nl 
  5004.771 &   3.11688 &  0.350 &  0  \nl 
  5017.920 &   3.12770 &  0.045 &  2  \nl 
  5020.106 &   3.12950 &  0.199 &  3  \nl 
  5157.707 &   3.24269 &  0.045 &  0  \nl 
  5166.664 &   3.25005 &  0.139 &  2  \nl 
  5170.188 &   3.25295 &  0.226 &  0  \nl 
  5187.813 &   3.26745 &  0.113 &  0  \nl 
  5201.190 &   3.27846 &  0.038 &  0  \nl 
  5208.194 &   3.28422 &  0.272 &  2  \nl 
  5211.599 &   3.28702 &  0.298 &  0  \nl 
  5212.832 &   3.28803 &  0.097 &  0  \nl 
  5237.367 &   3.30821 &  0.382 &  0  \nl 
  5241.231 &   3.31139 &  0.220 &  2  \nl 
  5249.536$^a$ &   3.31822 &  0.080 &  1  \nl 
  5250.509 &   3.31903 &  0.192 &  0  \nl 
  5279.110 &   3.34255 &  0.074 &  0  \nl 
  5286.269 &   3.34844 &  0.314 &  0  \nl 
  5291.105 &   3.35242 &  0.379 &  0  \nl 
  5315.959 &   3.37286 &  0.341 &  2  \nl 
  5319.197 &   3.37553 &  0.048 &  2  \nl 
  5322.354 &   3.37812 &  0.199 &  0  \nl 
  5341.464 &   3.39384 &  0.113 &  1  \nl 
  5370.738 &   3.41792 &  0.146 &  1  \nl 
  5387.182 &   3.43145 &  0.285 &  2  \nl 
  5395.701 &   3.43846 &  0.341 &  3  \nl 
  5411.299$^a$ &   3.45129 &  0.053 &  3  \nl 
  5423.721 &   3.46151 &  0.286 &  0  \nl 
  5432.508 &   3.46873 &  0.163 &  2  \nl 
  5439.274 &   3.47430 &  0.066 &  0  \nl 
  5484.907 &   3.51184 &  0.124 &  0  \nl 
  5494.495 &   3.51973 &  0.085 &  0  \nl 
  5509.467 &   3.53204 &  0.133 &  2  \nl 
  5554.169 &   3.56881 &  0.238 &  0  \nl 
  5561.464 &   3.57481 &  0.155 &  0  \nl 
  5562.494 &   3.57566 &  0.243 &  0  \nl 
  5585.622 &   3.59469 &  0.248 &  1  \nl 
  5601.092 &   3.60741 &  0.300 &  0  \nl 
  5617.280 &   3.62073 &  0.085 &  2  \nl 
  5620.783$^a$ &   3.62361 &  0.082 &  3  \nl 
  5621.720$^a$ &   3.62438 &  0.088 &  3  \nl 
  5627.097 &   3.62880 &  0.151 &  3  \nl 
\cutinhead{Q 1425+6039}
  4286.917 &   2.52638 &  0.041 &  2  \nl 
  4295.367 &   2.53333 &  0.372 &  0  \nl 
  4318.305 &   2.55220 &  0.212 &  0  \nl 
  4329.535 &   2.56144 &  0.241 &  0  \nl 
  4339.103 &   2.56931 &  0.213 &  0  \nl 
  4412.140 &   2.62939 &  0.362 &  0  \nl 
  4419.553 &   2.63549 &  0.117 &  0  \nl 
  4421.911 &   2.63743 &  0.195 &  0  \nl 
  4437.250 &   2.65004 &  0.242 &  0  \nl 
  4452.923 &   2.66294 &  0.211 &  1  \nl 
  4486.830 &   2.69083 &  0.145 &  1  \nl 
  4519.021 &   2.71731 &  0.311 &  3  \nl 
  4534.893 &   2.73037 &  0.368 &  0  \nl 
  4562.383 &   2.75298 &  0.057 &  3  \nl 
  4629.492 &   2.80818 &  0.060 &  1  \nl 
  4663.989 &   2.83656 &  0.084 &  1  \nl 
  4664.949 &   2.83735 &  0.230 &  1  \nl 
  4671.281 &   2.84256 &  0.111 &  1  \nl 
  4684.343 &   2.85330 &  0.088 &  0  \nl 
  4735.381 &   2.89528 &  0.254 &  0  \nl 
  4736.249 &   2.89600 &  0.056 &  0  \nl 
  4739.185 &   2.89841 &  0.211 &  2  \nl 
  4740.526 &   2.89952 &  0.338 &  1  \nl 
  4744.760 &   2.90300 &  0.305 &  2  \nl 
  4767.407 &   2.92163 &  0.113 &  0  \nl 
  4774.520 &   2.92748 &  0.127 &  0  \nl 
  4796.577 &   2.94562 &  0.128 &  0  \nl 
  4798.067 &   2.94685 &  0.232 &  0  \nl 
  4806.079 &   2.95344 &  0.277 &  0  \nl 
  4808.011 &   2.95503 &  0.236 &  1  \nl 
  4907.598 &   3.03695 &  0.271 &  2  \nl 
  4911.555 &   3.04020 &  0.260 &  1  \nl 
  4923.173 &   3.04976 &  0.344 &  3  \nl 
  4929.528 &   3.05499 &  0.107 &  1  \nl 
  4947.206 &   3.06953 &  0.359 &  1  \nl 
  4951.747 &   3.07327 &  0.216 &  3  \nl 
  4952.611$^a$ &   3.07398 &  0.160 &  3  \nl 
  4971.342 &   3.08938 &  0.222 &  3  \nl 
  4999.012 &   3.11214 &  0.223 &  0  \nl 
  5023.327 &   3.13215 &  0.079 &  1  \nl 
\cutinhead{Q 1442+2931}
  3903.794 &   2.21123 &  0.285 &  0  \nl 
  3905.104 &   2.21231 &  0.356 &  0  \nl 
  3971.901 &   2.26725 &  0.273 &  0  \nl 
  3973.539 &   2.26860 &  0.350 &  0  \nl 
  3978.611 &   2.27277 &  0.320 &  2  \nl 
  3999.285 &   2.28978 &  0.367 &  0  \nl 
  4032.404 &   2.31702 &  0.261 &  0  \nl 
  4066.457 &   2.34503 &  0.103 &  0  \nl 
  4075.212 &   2.35224 &  0.387 &  0  \nl 
  4113.835 &   2.38401 &  0.169 &  2  \nl 
  4154.397 &   2.41737 &  0.362 &  0  \nl 
  4159.561 &   2.42162 &  0.092 &  2  \nl 
  4197.781 &   2.45306 &  0.123 &  0  \nl 
  4199.317 &   2.45432 &  0.193 &  0  \nl 
  4202.369 &   2.45683 &  0.274 &  0  \nl 
  4210.112 &   2.46320 &  0.088 &  0  \nl 
  4222.104 &   2.47307 &  0.120 &  3  \nl 
  4225.344 &   2.47573 &  0.349 &  0  \nl 
  4230.729 &   2.48016 &  0.386 &  1  \nl 
  4234.958$^a$ &   2.48364 &  0.383 &  3  \nl 
  4237.895 &   2.48606 &  0.105 &  0  \nl 
  4242.205 &   2.48960 &  0.071 &  1  \nl 
  4248.379 &   2.49468 &  0.261 &  0  \nl 
  4273.131 &   2.51504 &  0.266 &  0  \nl 
  4308.993 &   2.54454 &  0.245 &  0  \nl 
  4318.667 &   2.55250 &  0.246 &  0  \nl 
  4323.667 &   2.55661 &  0.133 &  0  \nl 
  4341.328 &   2.57114 &  0.233 &  1  \nl 
  4348.721 &   2.57722 &  0.125 &  0  \nl 
  4367.699 &   2.59283 &  0.198 &  0  \nl 
  4378.564 &   2.60177 &  0.389 &  0  \nl 
  4390.083 &   2.61125 &  0.316 &  2  \nl 
\cutinhead{Q 1700+6416}
  4055.872 &   2.33633 &  0.191 &  0  \nl 
  4071.267 &   2.34899 &  0.092 &  1  \nl 
  4079.972 &   2.35615 &  0.191 &  0  \nl 
  4093.271 &   2.36709 &  0.029 &  1  \nl 
  4103.101 &   2.37518 &  0.237 &  1  \nl 
  4127.951 &   2.39562 &  0.054 &  1  \nl 
  4140.000 &   2.40553 &  0.286 &  0  \nl 
  4170.336 &   2.43048 &  0.325 &  0  \nl 
  4185.509 &   2.44297 &  0.296 &  1  \nl 
  4216.740 &   2.46866 &  0.299 &  0  \nl 
  4218.814 &   2.47036 &  0.212 &  0  \nl 
  4220.503 &   2.47175 &  0.161 &  1  \nl 
  4233.472 &   2.48242 &  0.328 &  0  \nl 
  4262.720 &   2.50648 &  0.160 &  0  \nl 
  4267.122 &   2.51010 &  0.115 &  0  \nl 
  4282.338 &   2.52262 &  0.243 &  0  \nl 
  4336.106 &   2.56684 &  0.287 &  0  \nl 
  4367.674 &   2.59281 &  0.106 &  2  \nl 
  4370.138 &   2.59484 &  0.147 &  1  \nl 
  4374.728 &   2.59861 &  0.330 &  0  \nl 
  4381.077 &   2.60384 &  0.045 &  1  \nl 
  4401.184 &   2.62038 &  0.293 &  0  \nl 
  4422.048 &   2.63754 &  0.204 &  0  \nl 
  4430.501 &   2.64449 &  0.208 &  0  \nl 
  4432.457 &   2.64610 &  0.235 &  0  \nl 
  4433.164 &   2.64668 &  0.118 &  0  \nl 
  4436.012 &   2.64903 &  0.264 &  0  \nl 
  4438.243 &   2.65086 &  0.269 &  0  \nl 
  4475.114 &   2.68119 &  0.168 &  0  \nl 
  4484.028 &   2.68852 &  0.291 &  0  \nl 
  4500.091 &   2.70174 &  0.077 &  0  \nl 
  4510.327 &   2.71016 &  0.047 &  2  \nl 
  4527.784 &   2.72452 &  0.275 &  0  \nl 
  4537.776 &   2.73274 &  0.322 &  0  \nl 
  4551.758 &   2.74424 &  0.180 &  0  \nl 
\cutinhead{Q 1946+7658}
  4215.363 &   2.46752 &  0.326 &  1  \nl 
  4216.741 &   2.46866 &  0.316 &  0  \nl 
  4233.834 &   2.48272 &  0.159 &  1  \nl 
  4254.062 &   2.49936 &  0.325 &  3  \nl 
  4272.777 &   2.51475 &  0.087 &  0  \nl 
  4310.893 &   2.54610 &  0.335 &  0  \nl 
  4322.006 &   2.55525 &  0.395 &  0  \nl 
  4345.969 &   2.57496 &  0.070 &  0  \nl 
  4348.446 &   2.57700 &  0.190 &  0  \nl 
  4354.806 &   2.58223 &  0.039 &  0  \nl 
  4358.269 &   2.58508 &  0.200 &  0  \nl 
  4381.860 &   2.60448 &  0.269 &  2  \nl 
  4388.862 &   2.61024 &  0.313 &  1  \nl 
  4391.371 &   2.61230 &  0.337 &  0  \nl 
  4397.191 &   2.61709 &  0.121 &  0  \nl 
  4403.181 &   2.62202 &  0.186 &  0  \nl 
  4406.600 &   2.62483 &  0.330 &  1  \nl 
  4421.906 &   2.63742 &  0.127 &  2  \nl 
  4425.030 &   2.63999 &  0.051 &  2  \nl 
  4439.273 &   2.65171 &  0.145 &  0  \nl 
  4445.566 &   2.65688 &  0.085 &  0  \nl 
  4460.339 &   2.66904 &  0.050 &  1  \nl 
  4464.265 &   2.67227 &  0.178 &  0  \nl 
  4489.117 &   2.69271 &  0.228 &  0  \nl 
  4500.549 &   2.70211 &  0.200 &  0  \nl 
  4506.008 &   2.70660 &  0.139 &  0  \nl 
  4531.280 &   2.72739 &  0.367 &  0  \nl 
  4537.618 &   2.73261 &  0.304 &  1  \nl 
  4570.602 &   2.75974 &  0.060 &  0  \nl 
  4578.117 &   2.76592 &  0.306 &  0  \nl 
  4593.453 &   2.77854 &  0.099 &  0  \nl 
  4619.984 &   2.80036 &  0.244 &  1  \nl 
  4633.098 &   2.81115 &  0.367 &  0  \nl 
  4647.124 &   2.82269 &  0.173 &  0  \nl 
  4656.047 &   2.83002 &  0.199 &  1  \nl 
  4692.409 &   2.85994 &  0.080 &  0  \nl 
  4749.612 &   2.90699 &  0.306 &  2  \nl 
  4753.097 &   2.90986 &  0.107 &  2  \nl 
  4759.734 &   2.91532 &  0.291 &  1  \nl 
  4780.079 &   2.93205 &  0.091 &  0  \nl 
  4818.925 &   2.96401 &  0.387 &  0  \nl 
  4836.081 &   2.97812 &  0.177 &  2  \nl 
  4842.774 &   2.98363 &  0.127 &  0  \nl 
  4846.812 &   2.98695 &  0.273 &  2  \nl 
  4871.646 &   3.00738 &  0.184 &  0  \nl 
  4881.021 &   3.01509 &  0.235 &  0  \nl 
  4901.713 &   3.03211 &  0.245 &  2  \nl 
  4904.279$^a$ &   3.03422 &  0.090 &  3  \nl 
\cutinhead{Q 2343+1232}
  3965.490 &   2.26198 &  0.084 &  1  \nl 
  3982.626 &   2.27607 &  0.065 &  0  \nl 
  3983.400 &   2.27671 &  0.113 &  0  \nl 
  3993.207 &   2.28478 &  0.051 &  2  \nl 
  4015.365 &   2.30300 &  0.129 &  0  \nl 
  4019.717 &   2.30659 &  0.278 &  0  \nl 
  4071.661 &   2.34931 &  0.150 &  0  \nl 
  4112.756 &   2.38312 &  0.310 &  1  \nl 
  4150.644 &   2.41428 &  0.147 &  0  \nl 
  4195.085 &   2.45084 &  0.331 &  2  \nl 
  4207.765 &   2.46127 &  0.056 &  0  \nl 
  4209.998 &   2.46311 &  0.189 &  2  \nl 
  4210.993 &   2.46393 &  0.104 &  0  \nl 
  4236.032 &   2.48452 &  0.373 &  0  \nl 
  4237.870 &   2.48604 &  0.375 &  0  \nl 
  4239.561 &   2.48743 &  0.198 &  2  \nl 
  4248.781 &   2.49501 &  0.305 &  0  \nl 
  4257.533$^a$ &   2.50221 &  0.106 &  1  \nl 
\enddata

\tablenotetext{a}{This system shows detectable \civ absorption.}

\end{planotable}

\clearpage

\begin{planotable}{cccccccc}
\tablewidth{0pc}
\tablecaption{Ly$\alpha$ Systems with Metals}
\tablehead{
\colhead{QSO}              &\colhead{$z_{abs}$}    
 &\colhead{N(\textsc{H~i})$^a$}    &\colhead{N(\textsc{C~iv})$^a$}
 &\colhead{log T(\textsc{H~i})$^b$}    &\colhead{log T(\textsc{C~iv})$^b$} 
 &\colhead{[C/H]$_{\rm P}^c$}     &\colhead{[C/H]$_{\rm C}^d$} 
         }
\startdata
\cutinhead{Associated Systems$^e$}
1107$+$4847 &2.95614 &13.99 &12.09 &$<4.6$ &$<5.0$ &$>-1.85$ &$>-2.76$ \nl
1422$+$2309 &3.58914$^f$ &13.25 &12.55 &$<4.2$ &$<5.3$ &$>-0.67$ &$>-1.56$ \nl
   &3.62361 &13.78 &13.45$^g$ &$<4.4$ &$<4.9^g$ &$>-0.28$ &$>-1.19$ \nl
   &3.62438 &13.99$^h$ &13.59$^h$ &...$^h$ &...$^h$ &$>-0.35$  &$>-1.26$\nl
1425$+$6039&3.15341$^f$&14.40$^j$&13.06$^j$&...$^j$&$<5.8$&$>-1.29$&$>-2.20$\nl
            &3.15401$^f$ &...$^j$   &...  $^j$ &...$^j$ &$<5.5$ &... &... \nl
1946$+$7658 &3.03422 &14.06$^k$ &12.43&...$^k$&$<5.2$ &$>-1.58$ &$>-2.49$\nl
2343$+$1232 &2.56892$^f$ &13.49 &13.47 &$<4.4$ &$<4.6$ &$>+0.03$ &$>-0.88$ \nl
    &2.56940$^f$ &13.58 &13.36$^l$ &$<4.3$ &$<4.6^l$ &$>-0.17$ &$>-1.08$\nl
\cutinhead{Intervening Systems$^e$}
1107$+$4847 &2.72224 &13.74 &12.98 &$<4.3$ &$<4.4$ &$>-1.10$ &$>-1.62$ \nl
1422$+$2309 &3.13247$^f$ &13.71 &12.47 &$<4.9$ &$<5.8$ &$>-1.58$ &$>-2.10$ \nl
            &3.31822 &13.96 &12.47 &$<4.7$ &$<5.9$ &$>-1.83$ &$>-2.35$ \nl
            &3.45129 &13.99 &12.69 &$<4.6$ &$<5.5$ &$>-1.64$ &$>-2.16$ \nl
1425$+$6039&3.07398&13.66&12.44$^i$ &$<4.4$ &$<4.6^i$ &$>-1.56$ &$>-2.08$ \nl
1442$+$2931 &2.48364 &13.30 &12.75 &$<4.2$ &$<4.3$ &$>-0.89$ &$>-1.41$\nl
2343$+$1232 &2.50221 &13.78 &11.95 &$<4.4$ &$<5.2$ &$>-2.17$ &$>-2.69$ \nl
\enddata
\end{planotable}

\noindent {\bf Notes to Table 3}

\noindent $^a$   {Column density (logarithmic) of
\textsc{H~i} and \textsc{C~iv} based on Voigt profile
fits to the \lya line and the \civ doublet.}

\noindent $^b$   Temperature of the gas implied by the width
of the \lya and the \civ absorption lines.

\noindent $^c$   {Metallicity of the gas assuming the gas is photoionized.
For intervening systems the photoionization results using the
$\alpha=-1.8$ 
Haardt \& Madau (1996) quasar spectrum including intervening opacity
are adopted. For associated systems the photoionization results using
the unabsorbed $\alpha=-1.8$ quasar spectrum of 
Haardt \& Madau (1996) are used.}

\noindent $^d$   {Metallicity of the gas assuming the gas is
collisionally ionized and is in thermal and ionization equilibria.}

\noindent $^e$   {Type of the absorption system.
All intervening systems are more than
10,000 \kms away from the emission redshift of the quasars, and
all associated systems are within 3000 \kms of the quasar emission
redshifts.}

\noindent $^f$   {This system is within 50 \kms of another strong
\lya absorption and and would otherwise not be included in the
\lya sample in Table 2.}

\noindent $^g$   {The \civ absorption at this redshift has two components.
The values given are for the stronger component. The weaker component
adds only 18\% to the \civ column density.}

\noindent $^h$   {Both the \lya and \civ absorption require two 
components of comparable column densities to fit. The values given
are for the sum of the two.}

\noindent $^i$   {The \civ $\lambda$1548 absorption is blended with
the \civ $\lambda$1550 absorption at $z=3.06702$. The values are obtained
after deblending.}

\noindent $^j$   {The \lya lines of these two systems are very close
to each other and require 3 components to fit. It is impossible to
determine the \nhi\ of each system. The values listed are for the sum
of the two systems (3 components).}

\noindent $^k$   {The \lya line requires two closely spaced 
components of comparable column densities
to fit. The values given are for the sum of the two.
The formal $b$ values of the components are 13.5 and 13.9 \kms,
implying log T$<4.5$ and 4.8 respectively.}

\noindent $^l$   {The \civ absorption at this redshift has two components.
The values given are for the stronger component. The weaker component
adds only 30\% to the \civ column density.}

\clearpage

\begin{planotable}{cccccccc}
\tablewidth{0pc}
\tablecaption{Results based on Composite Spectra}
\tablehead{
\colhead{Sample$^a$}      &\colhead{No. Systems$^b$}   &\colhead{$<z>^b$}  
 &\colhead{S/N$^c$}       &\colhead{log $<$N(\textsc{H~i})$>^d$}
 &\colhead{$w_r(1548)^e$} &\colhead{log $<$N(\textsc{C~iv})$>^f$} 
 &\colhead{[C/H]$^g$}     }
\startdata
A  &282  &2.748 &1860  &13.75       &$<0.088$ m\AA\  &$<10.34$  &$<-3.49$  \nl
B  &156  &2.727 &1390  &13.63       &$<0.113$ m\AA\  &$<10.45$  &$<-3.15$  \nl
C  &126  &2.774 &1240  &13.88       &$<0.134$ m\AA\  &$<10.52$  &$<-3.55$  \nl
D  &153  &2.526 &1300  &13.75       &$<0.120$ m\AA\  &$<10.47$  &$<-3.36$  \nl
E  &129  &3.012 &1330  &13.75       &$<0.123$ m\AA\  &$<10.48$  &$<-3.35$  \nl
F  & 27  &2.846 & 600  &13.75       &$<0.251$ m\AA\  &$<10.79$  &$<-3.04$  \nl
G  & 12  &3.030 & 380  &13.75       &$<0.499$ m\AA\  &$<11.09$  &$<-2.74$  \nl
\enddata

\tablenotetext{a}{Samples are defined in section 3.3. }

\tablenotetext{b}{Number of Ly$\alpha$ systems used to form the composite
spectrum and the mean redshift of the sample.}

\tablenotetext{c}{The S/N per 6.6 km/s resolution element near the
\civ $\lambda$1548 line in the composite spectrum, calculated from
the formal error spectrum.}

\tablenotetext{d}{Mean \textsc{H~i} column density of the sample
estimated from the column density distribution $f(N)\propto N^{-1.5}$.}

\tablenotetext{e}{Upper limit on the rest-frame equivalent width
of \civ $\lambda$1548 at the 95.5\% confidence level based on the
results of Monte Carlo simulations.}

\tablenotetext{f}{Upper limit on N(\textsc{C~iv}) at the 95.5\%
confidence level, derived from the equivalent width limit in
column 5 assuming linear part of curve of growth.}

\tablenotetext{g}{Upper limit on [C/H] at the 95.5\% confidence
level; see section 3.2 for details.}

\end{planotable}

\clearpage

{\bf Figure 1} - Distribution of \lya clouds in the spectrum
of Q 1946+7658 that are included in our sample of \lya systems based
on selection by optical depth (Table 2).  The actual N(\textsc{H~i}) of
the clouds are from Voigt profile fitting by Kirkman \& Tytler (1997).
The dashed line, scaled arbitrarily to roughly match the observed
distribution,  indicates the expected distribution from the known
column density distribution of $f(N)\propto N^{-1.5}$. It is seen that
the \nhi\ of most of the \lya lines selected falls within the
desired range of \nrange.

{\bf Figure 2} - Profiles of \lya and \civ $\lambda$1548, 1550 in the 
16 \lya systems with detectable \civ absorption (Table 3). Note
that 6 of the systems form close redshift pairs and are displayed
in 3 panels. The type of the absorption systems (Table 3) are indicated:
``I'' for intervening and ``A'' for associated.
The profiles are shown in velocity space in the rest frame
of the absorbers. 

{\bf Figure 3} - Composite spectrum for Sample A. The spectrum is shown
in velocity space relative to \civ $\lambda$1548.20 \AA. The
velocity bin size is 3.3 \kms. The two vertical lines mark the
expected position of the \civ $\lambda\lambda$1548.20, 1550.77
absorption. The smooth curve is the fitted continuum, which
includes the absorption by a \civ doublet with $b=10$ \kms 
and with $w_r(1548)$ equal to the 4$\sigma$ upper limit.

{\bf Figure 4} - Metallicity as a function of \textsc{H~i} column density for
quasar absorption line systems. The box labeled ``DLA'' is for damped \lya
systems using [Fe/H] as the metallicity indicator (Lu et al 1996).
The box outlined with dotted lines indicates the metallicity distribution
of damped \lya systems if [Zn/H] is used as a metallicity indicator
instead (Pettini et al 1997). The box labeled ``LLS'' is for Lyman limit
systems studied by Steidel (1990). The box labeled ``\lya'' is for \lya
clouds with \nhi$>10^{14.5}$ \cm2 studied by Cowie et al (1995), Tytler
et al (1995), Songaila \& Cowie (1996) and Rauch et al (1997). The sizes
of the boxes roughly indicate the spread in the metallicity distribution
(vertical direction) and the range of \nhi\ (horizontal direction). For
the \lya clouds, the spread in metallicity is not well documented in any
of the references quoted; a spread of 1 dex is assumed. The upper limit
at the lowest column density is from this study.

{\bf Figure 5} - Constraints on the metallicity distribution of \lya clouds.
The meaning of this figure is explained in section 4.2.

\clearpage

\begin{figure}
\plotone{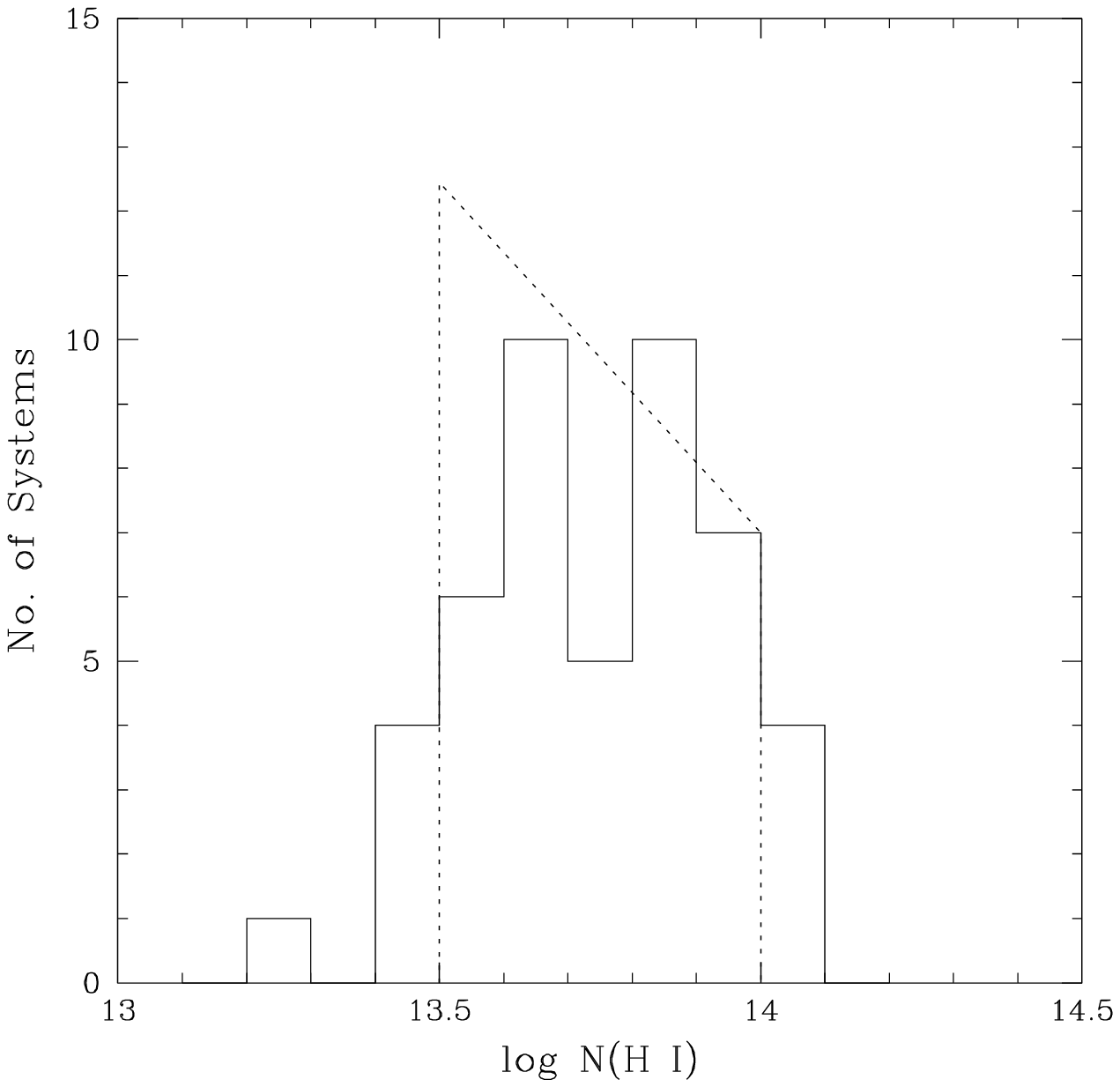}
\figcaption{}
\end{figure}

\begin{figure}
\plotone{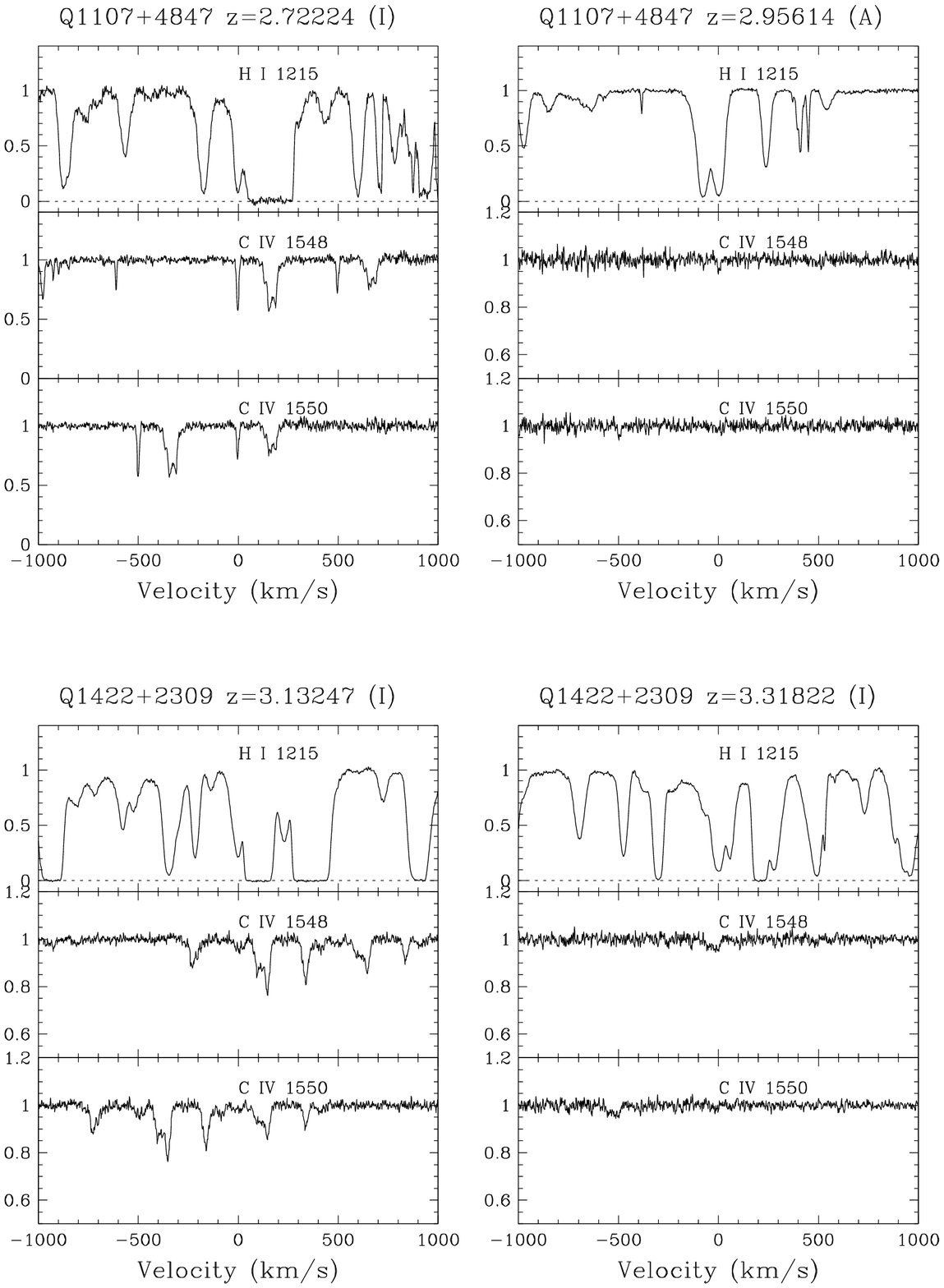}
\figcaption{}
\end{figure}

\begin{figure}
\plotone{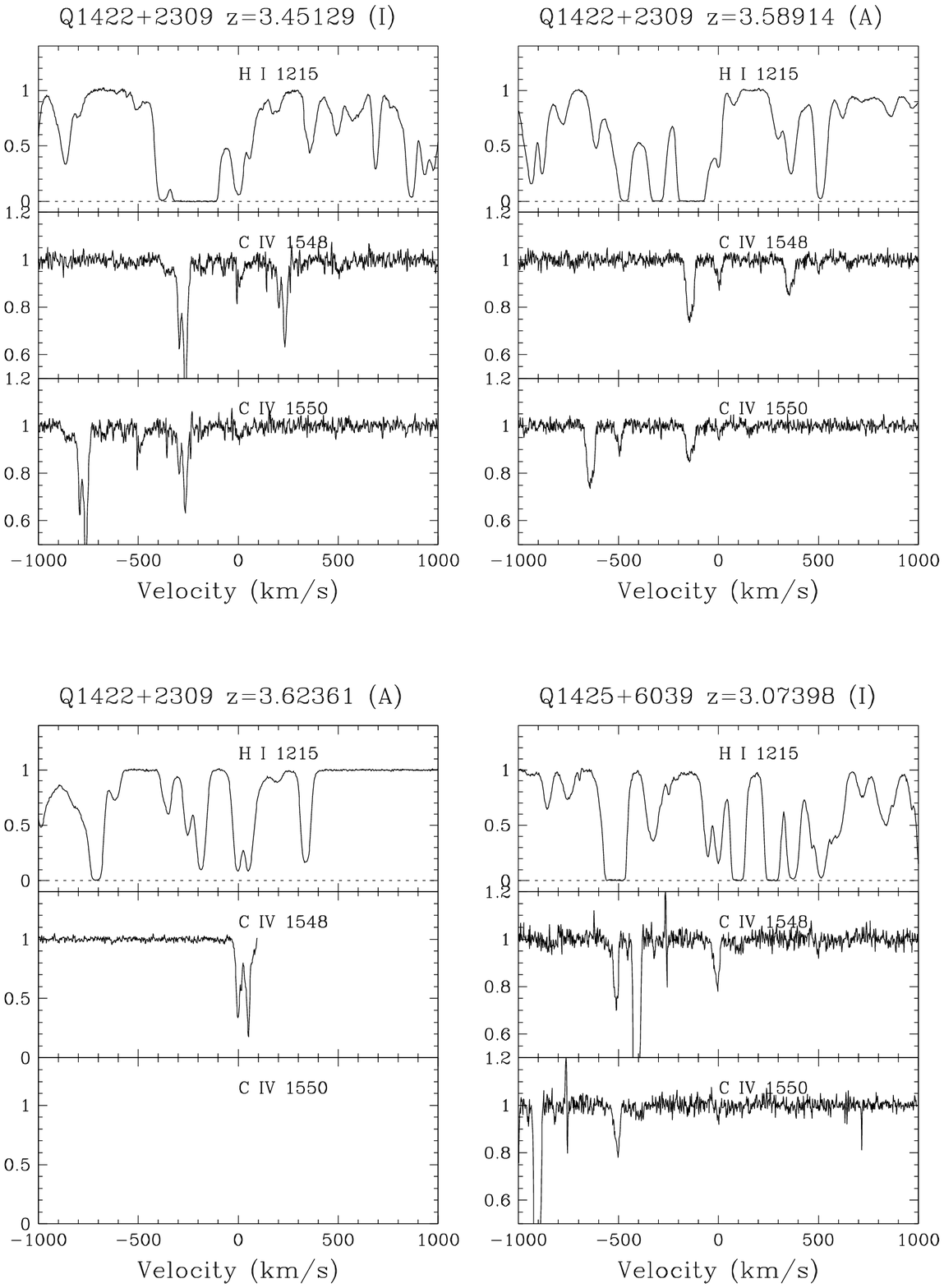}
\end{figure}

\begin{figure}
\plotone{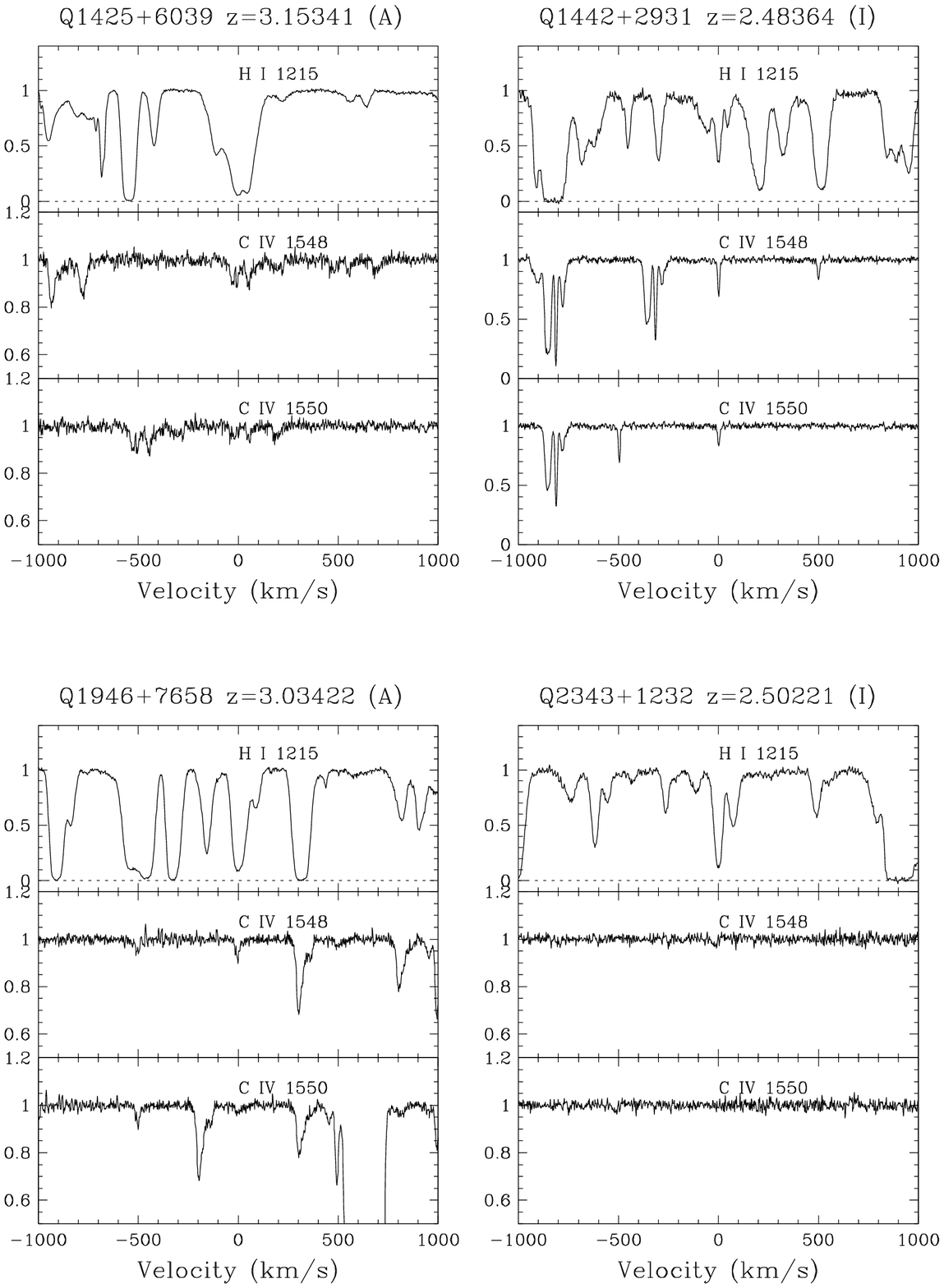} 
\end{figure}

\begin{figure}
\plotone{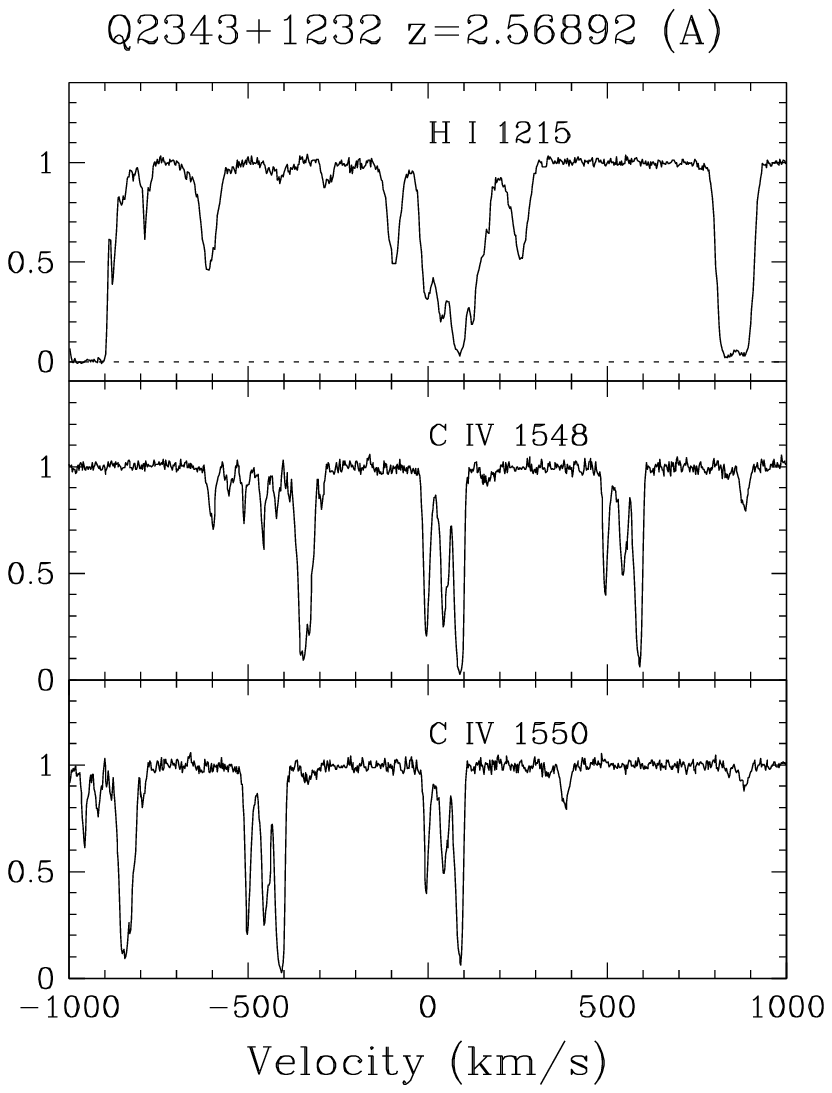}
\end{figure}

\begin{figure}
\plotone{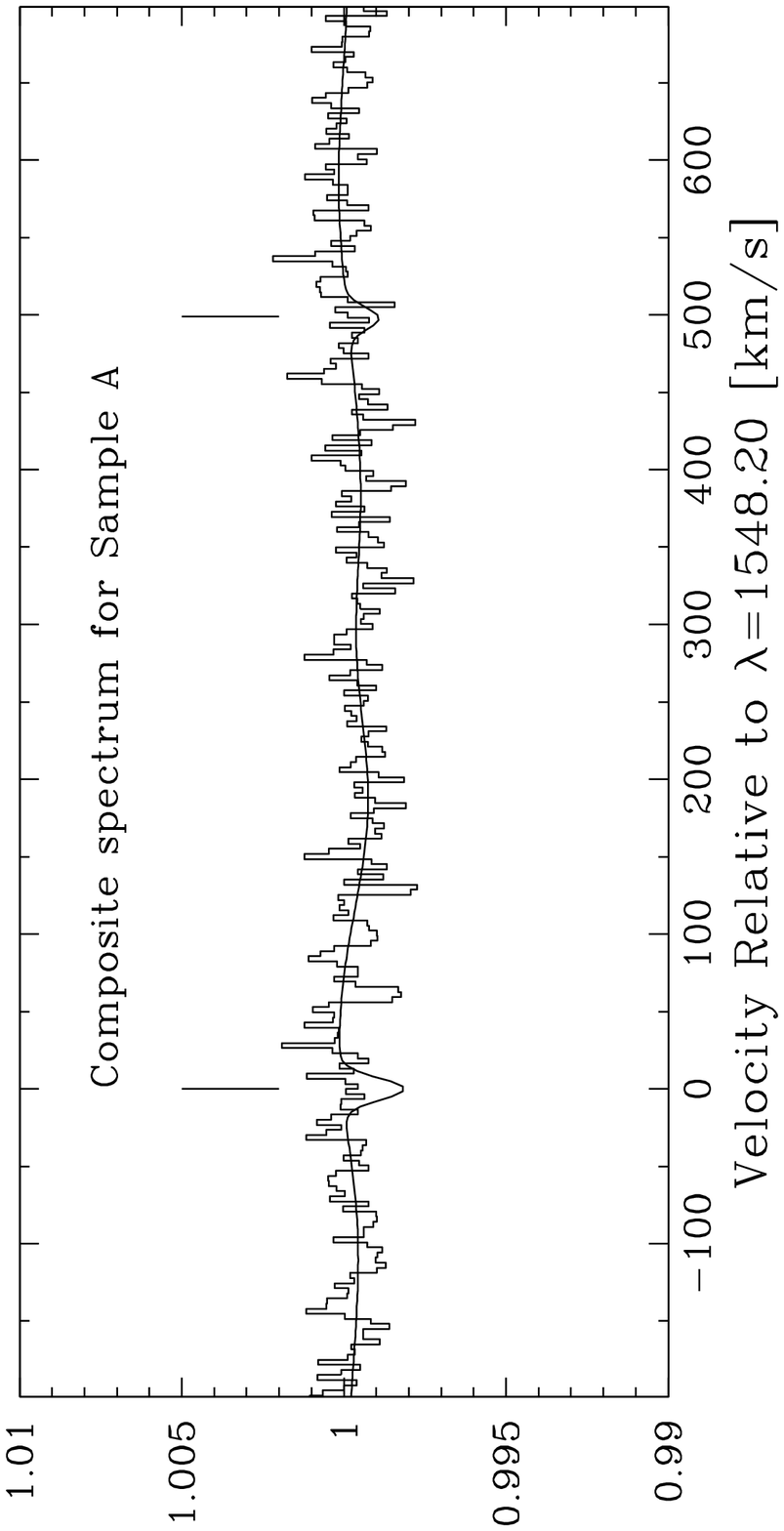}
\figcaption{}
\end{figure}

\begin{figure}
\plotone{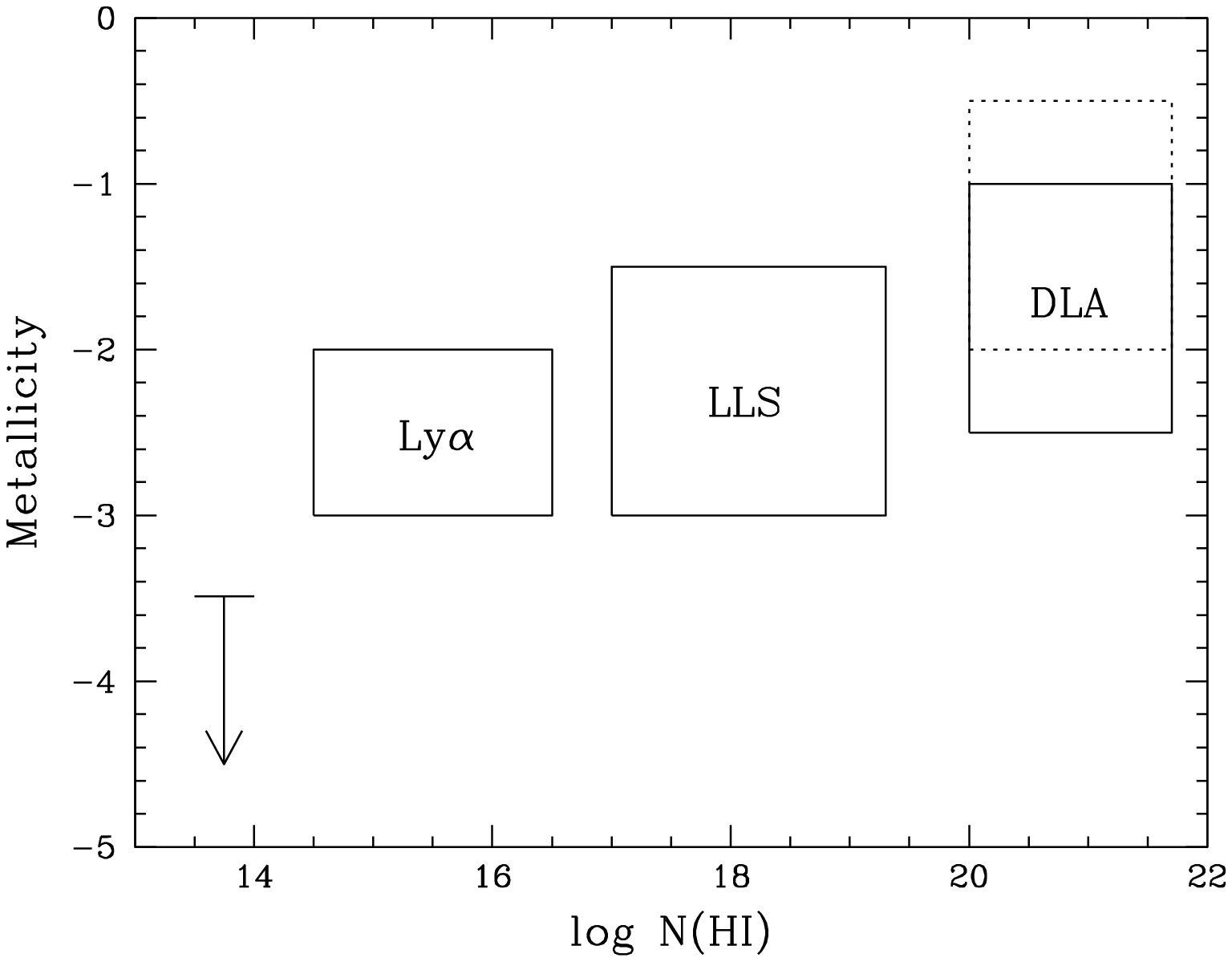}
\figcaption{}
\end{figure}

\begin{figure}
\plotone{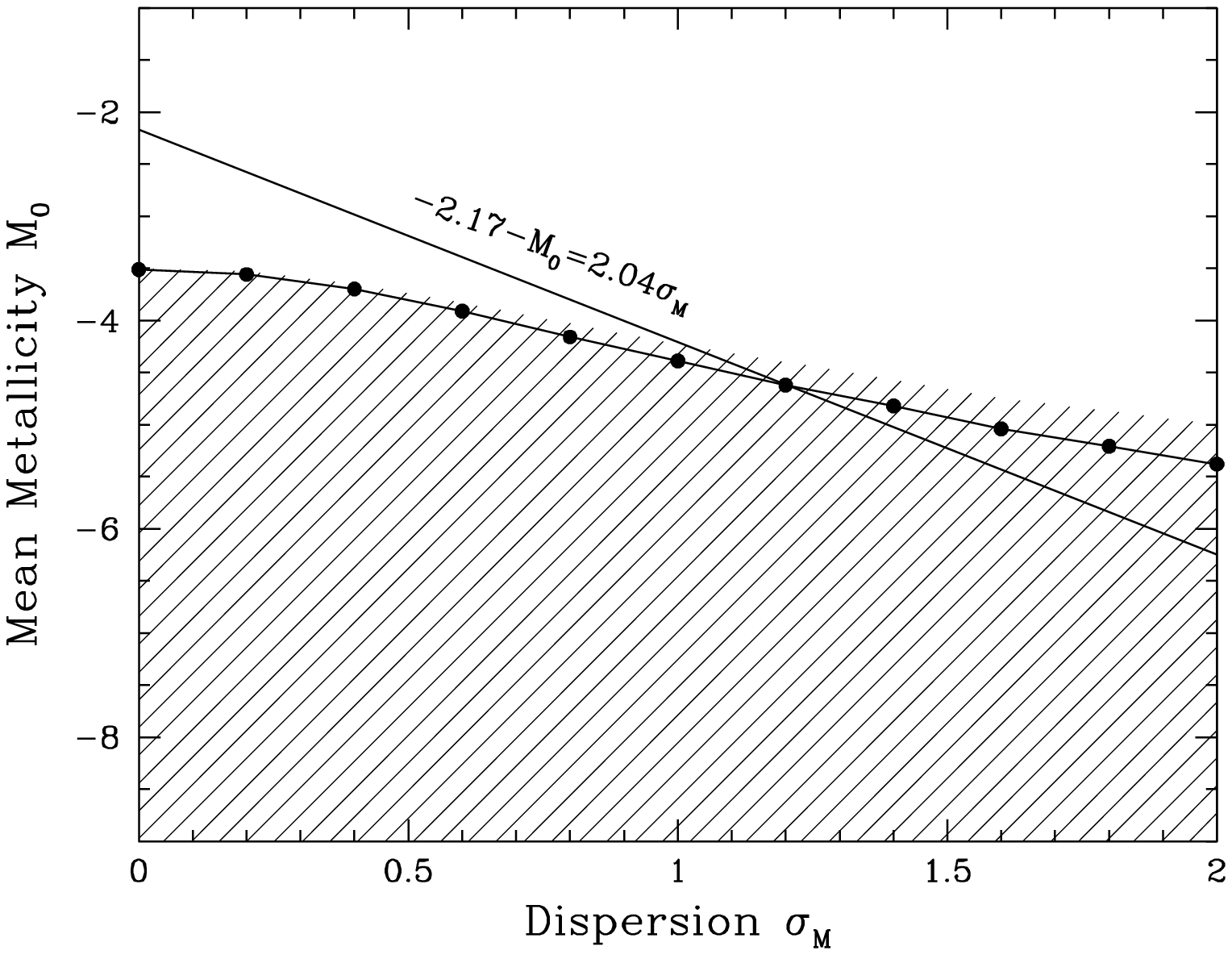}
\figcaption{}
\end{figure}

\end{document}